\documentclass[aps,prd,nofootinbib,superscriptaddress,preprint,eqsecnum,showkeys,showpacs,preprintnumbers]{revtex4-2}
\usepackage{amsmath}
\usepackage{amssymb}
\usepackage{graphicx}
\usepackage{epsfig}
\usepackage{xcolor,color}
\usepackage{url}
\usepackage{bm}
\usepackage{mathrsfs}
\usepackage[utf8]{inputenc}
\usepackage{hyperref}
\usepackage{enumerate}
\usepackage{amsthm}
\usepackage{bbm}
\usepackage[normalem]{ulem}
\usepackage{upgreek}
\usepackage{tensor}
\usepackage{siunitx}
\usepackage{orcidlink}
\usepackage[caption=false]{subfig}
\usepackage{colortbl}
\usepackage{hyperref} 
\usepackage{appendix} 
\usepackage{cleveref}
\usepackage{multirow}
\usepackage{booktabs}
\usepackage{threeparttable}
\usepackage[export]{adjustbox}
\usepackage{times}
\usepackage{commath}
 
 \allowdisplaybreaks[3]
\hypersetup{colorlinks=true,linkcolor=blue,anchorcolor=blue,citecolor=blue}  
\definecolor{colour1}{HTML}{0571b0} 
\definecolor{colour2}{HTML}{92c5de} 
\definecolor{colour3}{HTML}{f4a582} 
\definecolor{colour4}{HTML}{ca0020} 
\definecolor{colour5}{HTML}{fe4a49} 
\definecolor{colour6}{HTML}{2d3092} 

\hypersetup{colorlinks=true, linkcolor=colour6, citecolor=colour6,
filecolor=colour6, urlcolor=colour6}

\theoremstyle{plain}

\DeclareMathOperator{\arcsinh}{arcsinh}

\newcommand{\bea}{\begin{eqnarray*}}
\newcommand{\eea}{\end{eqnarray*}}
\newcommand{\bean}{\begin{eqnarray}}
\newcommand{\eean}{\end{eqnarray}}

\newcommand{\vvr}{\mbox{\boldmath${r}$}}
\newcommand{\vp}{\mbox{\boldmath${p}$}}
\newcommand{\vP}{\mbox{\boldmath${P}$}}

\newcommand{\be}{\begin{equation}}
\newcommand{\ee}{\end{equation}}

\newcommand\beq{\begin{equation}}
\newcommand\eeq{\end{equation}}
\def\bea{\begin{eqnarray}}
\def\eea{\end{eqnarray}}

\begin{document}

\title{Effective one-body theory of spinless binary evolution dynamics}

\author{Jiliang {Jing},\orcidlink{0000-0002-2803-7900}\footnote{Corresponding author: jljing@hunnu.edu.cn} }
 \affiliation{Department of Physics, Key Laboratory of Low Dimensional Quantum Structures and Quantum Control of Ministry of Education, and Synergetic Innovation
Center for Quantum Effects and Applications, Hunan Normal
University, Changsha, Hunan 410081, P. R. China}
\affiliation{Center for Gravitation and Cosmology, College of Physical Science and Technology, Yangzhou University, Yangzhou 225009, P. R. China}

\author{Sheng Long}
 \affiliation{Department
of Physics, Key Laboratory of Low Dimensional Quantum Structures and
Quantum Control of Ministry of Education, and Synergetic Innovation
Center for Quantum Effects and Applications, Hunan Normal
University, Changsha, Hunan 410081, P. R. China}

\author{Weike Deng}
\affiliation{Department
of Physics, Key Laboratory of Low Dimensional Quantum Structures and
Quantum Control of Ministry of Education, and Synergetic Innovation
Center for Quantum Effects and Applications, Hunan Normal
University, Changsha, Hunan 410081, P. R. China}

\author{Jieci Wang\footnote{ jcwang@hunnu.edu.cn}} 
\affiliation{Department
of Physics, Key Laboratory of Low Dimensional Quantum Structures and
Quantum Control of Ministry of Education, and Synergetic Innovation
Center for Quantum Effects and Applications, Hunan Normal
University, Changsha, Hunan 410081, P. R. China}

\begin{abstract}
The effective one-body (EOB) theory  provides an innovative framework for analyzing the dynamics of binary systems, as articulated by Hamilton's equations. This paper investigates a self-consistent EOB theory specifically tailored for the dynamics of such systems. Our methodology begins by emphasizing how to effectively utilize the metrics derived from scattering angles in the analysis of binary black hole mergers. We then construct an effective Hamiltonian and formulate a decoupled, variable-separated Teukolsky-like equation for $\psi^B_4$. Furthermore, we present the formal solution to this equation, detailing the energy flux, radiation-reaction force (RRF), and waveforms for the ``plus" and ``cross" modes generated by spinless binaries. Finally, we carry out numerical calculations using the EOB theory and compare the results with numerical relativity (NR) data from the SXS collaboration. The results indicate that to the innermost stable circular orbit, the binding energy---angular momentum relation differs from the NR results by less than $5$\textperthousand, with a larger mass ratio yielding better agreement.
\end{abstract}


\maketitle

\newpage



\section{Introduction}
The successful detection of gravitational waves (GWs) ~\cite{Abbott2016,Abbott20162,Abbott2017,Abbott20172,Abbott20173,Abbott2019,Abbott20211,Abbott20212,Abbott20213} has been facilitated by significant advancements in experimental physics, data analysis, and theoretical modeling of sources. Such modeling is critical for estimating the physical parameters associated with the inspiral and coalescence of relativistic compact binaries. 
A crucial aspect of analyzing the evolution of binary systems is studying the dynamics associated with the inspiral, plunge, and coalescence phases. A novel approach to examining these dynamics is the EOB theory, which is grounded in the post-Newtonian (PN) approximation initially introduced by Buonanno and Damour \cite{Damour1999, Damour20002, Damour2007,Damour20091, Damour2000,Damour2001,DamourH,Pan2,Damour2009}. This theory effectively maps the two-body problem onto an EOB framework. EOB theory has proven to be a successful model for describing the gravitational radiation emitted by binary black holes and serves as the foundation for computing numerous gravitational waveform templates \cite{Taracchini,Bohe}  employed in the data analysis of GW signals \cite{Abbott2016,Abbott20162,Abbott2017,Abbott20172}.

To extend beyond the assumption that the ratio of velocity to the speed of light $ v/c$ remains small within the context of the EOB theory based on the 
PN 
approximation, Damour \cite{Damour2016} proposed an alternative EOB 
framework rooted in the post-Minkowskian (PM) approximation, which has garnered considerable attention  \cite{Damour2017, Damour2018,Damour2018new,Antonelli2019,Damour2019, Damour2020,HeLin2016, Jing2019,Blanchet2018, Cheung2018,Vines2019,Cristofoli2019, Collado2019, Bern2019,Bern20192, Plefka2019,Cheung2020}.

It is evident that a self-consistent EOB theory requires that all formulas and quantities within the system, such as the Hamiltonian, the stress-energy tensor, GW energy flux, the RRF, and the waveform, derive from a unified physical model. To determine the expressions of the RRF and waveform for the ``plus" and ``cross" modes of GWs, it is necessary to find the decoupled and variable-separable equation for  null tetrad component of the gravitational perturbed Weyl tensor $ \psi^B_{4} $ in the effective spacetime. Recently, we \cite{Jing1} derived the decoupled equations of $ \psi^{B}_{4} $ for even and odd parities in the Regge-Wheeler gauge \cite{Thompson} by separating the perturbation part of the metric into odd and even parities. However, calculations for this model are arduous because we must simultaneously solve two equations pertaining to odd and even parities. We \cite{Jing} also derived another decoupled and variable-separable equation for $ \psi^B_{4} $ in the effective spacetime by adopting a gauge in which $ \psi_{1}^{B} $ and $ \psi_{3}^{B} $ vanish. This is feasible because, in linear perturbation theory, $ \psi_{0}^{B} $ and $ \psi_{4}^{B} $ are gauge-invariant, while $ \psi_{1}^{B} $ and $ \psi_{3}^{B} $ are not \cite{Chandrasekhar}. We \cite{JingXZ} noted that in this gauge, the decoupled equation can only be separated between radial and angular variables in slowly rotating background spacetime. In this paper, we aim to construct a self-consistent EOB theory for binary systems by adopting a new gauge in which the decoupled equation can be separated variables for general case. It is imperative to emphasize that the metric derived in a previous study \cite{Jing3} was based on results related to scattering angles, yielding parameters applicable only to scattering states. To enable the effective application of the metric obtained from scattering angles in the analysis of binary black hole mergers, we must implement several adjustments.

Subsequently, we derive a decoupled and variable-separated Teukolsky-like equation for $ \psi^B_4 $, incorporating a source term characterized by the stress-energy tensor within the effective spacetime, and construct a formal solution to this equation. We elucidate the energy flux, RRF, and waveform for the ``plus" and ``cross" modes of GWs generated by spinless binary systems.

To test this self-consistent EOB theory, we present a comparison between the binding energy $ E_{b}(j) $ (where $ j $ represents orbital angular momentum) calculated using EOB theory and results from NR simulations, as binding energy is a critical component in computing gravitational waveforms.

The remainder of this paper is organized as follows: In Sec. II, we present the formulas for the Hamiltonian equations within EOB theory. In Sec. III, we emphasize the necessity of employing analytic continuation. In Sec. IV, we address the formal solution for the Teukolsky-like equation and present the energy flux, RRF, and waveform for the ``plus" and ``cross" modes of GWs. Section V carries out numerical  calculations using EOB theory based on the adiabatic approximation and compares the results with NR data. We conclude with a summary and discussion in the final section.


\section{Hamilton equations for EOB theory of spinless binaries}

For a real spinless two-body system, the basic idea of the EOB theory  is to map the dynamics of two compact objects with masses ($m_1$, $m_2$) into the dynamics of an effective test particle with mass $m_0=m_1m_2/(m_1+m_2)  $ orbits around a massive black hole characterized by an effective metric $g^{\text{eff}}_{\mu\nu}$ with mass parameter $M =m_1+m_2 $. 
The EOB dynamics are governed by the Hamilton equations  \cite{Damour2001,Taracchini1}
  \begin{align}\label{HEq}    
  \frac{d\vvr}{d\hat{t}}&=\Big\{\vvr,\hat{H}\big[g^{\text{eff}}_{\mu\nu}\big]\Big\}=\frac{\partial \hat{H}\big[g^{\text{eff}}_{\mu\nu}\big]}{\partial \vp}\,,\\
    \frac{d\vp}{d\hat{t}}&=\Big\{\vp,\hat{H}\big[g^{\text{eff}}_{\mu\nu}\big]\Big\}+\hat{\bm{\mathcal{F}}}\big[g^{\text{eff}}_{\mu\nu}\big] =-\frac{\partial \hat{H}\big[g^{\text{eff}}_{\mu\nu}\big]}{\partial \vvr}
    +\hat{\bm{\mathcal{F}}}\big[g^{\text{eff}}_{\mu\nu}\big]\,, \label{EOM3}
    \end{align}
where $\hat{t}\equiv t/ M $, $
\hat{H}[g^{\text{eff}}_{\mu\nu}]$ is the reduced EOB Hamiltonian \cite{BarausseH,BarausseH1,Barausse}, and  $\hat{\bm{\mathcal{F}}}[g^{\text{eff}}_{\mu\nu}]=\bm{\mathcal{F}}[g^{\text{eff}}_{\mu\nu}]/m_0$ is the reduced RRF.

By employing the energy relation  
$\mathcal{E}_0=\frac{\mathcal{E}^2-m_1^2-m_2^2}{2(m_1+m_2)}$ \cite{Jing,Jing1,JingXZ} between the relativistic energy $\mathcal{E}$ of real two-body system and the relativistic energy $\mathcal{E}_0$ of EOB system, we know that the improved reduced EOB Hamiltonian appearing in Eq. (\ref{HEq}) can be expressed as
  \begin{eqnarray}
\hat{H}\big[g^{\text{eff}}_{\mu\nu}\big]=\frac{1}{\nu} \sqrt{1+2\nu \left(\hat{H}_{\text{eff}} [g_{\mu\nu}^{\text{eff}}]-1 \right)}\,,\label{HHH}
  \end{eqnarray}
where $\hat{H}_{\text{eff}}\big [g_{\mu\nu}^{\text{eff}}\big]=\frac{1}{ \sqrt{g^{\text{eff}}_{tt}}} \,\sqrt{m_0^2 -g^{\text{eff}}_{ij}\,p_i\, p_j+ {\cal Q}_4(p)} $ is an effective Hamiltonian, ${\cal Q}_4(p)$ is a quartic term in the space momenta $p_i$ which was introduced in Ref.~\cite{Damour20002}.

It can be seen that the EOB theory includes three important components: effective metric, RRF, and waveform.  We will study them one by one.

\section{Effective metric  in the EOB theory  of spinless binaries}

In the EOB theory \cite{Damour1999}, the fundamental concept is to relate the two-body problem to an EOB problem through a systematic mapping process. This mapping can be articulated by sequentially identifying the scattering angles for the two systems. Utilizing the scattering angle  definition 
$
\chi=-\pi+2 J \int_{r_{\rm min}}^{\infty}\frac{d r}{r^2\sqrt{P_r^2}}, \label{chireal}
$
we previously identified an effective metric for spinless binaries \cite{Jing3}. However, it is imperative to emphasize that the metric constructed in reference \cite{Jing3} was derived directly from the results related to scattering angles, resulting in parameters applicable only to scattering states (i.e., $\gamma > 1$ or $p_\infty^2 > 0$). If this metric is applied directly to bound states (i.e., $\gamma < 1$ or $p_\infty^2 < 0$), it would yield complex-value results that do not correspond to physical phenomena. Consequently, adjustments must be made to appropriately study the merger of bound binary black holes, as discussed in the following.

Starting from Bern's expression of the conservative Hamiltonian \cite{Bern20192} for a relativistic massive spinless two-body system, we can derive the radial momentum as a function of the radial coordinate $r$, which, up to the 4PM order, is expressed as \cite{Bern20192, Jing3} 
 \begin{eqnarray}
 P_r^2 = \frac{P_0 r^2 - J^2}{r^2} + P_1 \Big(\frac{G}{r}\Big) + P_2 \Big(\frac{G}{r}\Big)^2 + P_3 \Big(\frac{G}{r}\Big)^3 + P_4 \Big(\frac{G}{r}\Big)^4,\nonumber \\
 \end{eqnarray}
  where $P_n$ (n=1, 2, 3, 4) denotes the coefficients in the PM expansion of the following Fourier transform of the scattering amplitude \cite{Bern20192} 
 \begin{eqnarray}
 \widetilde{\mathcal{M}}(r, E) &=& \frac{1}{2E} \int \frac{d^3 \mathbf{q}}{(2\pi)^3} \mathcal{M}(\mathbf{q}, p^2_\infty\nonumber \\  &=& p_\infty^2(E)) e^{-i \mathbf{q} \cdot \mathbf{r}} 
 = \sum_{n=1}^\infty P_n \left(\frac{G}{r}\right)^n \,.
 \label{FT}
  \end{eqnarray}
The coefficients $P_n$ are,  for clarity, presented in Appendix \ref{Pn}. 
For concise representation, we take  $
P_0 = \frac{(\mathcal{E}^2-(m_1-m_2)^2)(\mathcal{E}^2 - M^2)}{4\mathcal{E}^2} = \left(\frac{\mu}{\Gamma}\right)^2 p_\infty^2 = \left(\frac{\mu}{\Gamma}\right)^2 (\gamma^2 - 1),
$ with $M = m_1 + m_2$, $\Gamma = \frac{\mathcal{E}}{M}$, and $\mu = \frac{m_1 m_2}{M}$. Here, $p_\infty^2 = (\gamma^2 - 1)$, and $\mathcal{E}$ denotes the relativistic energy of the real two-body system. For scattering states, we have $\mathcal{E} - M > 0$, whereas for bound states, $\mathcal{E} - M < 0$, which shows that the primary distinction between these two scenarios  is that $p_\infty^2 > 0$ ($\gamma > 1$) characterizes scattering states, while $p_\infty^2 < 0$ ($\gamma < 1$) pertains to bound states.

\if
Then, by using the  definition 
\begin{align}
\chi^{\text{Nor}}=-\pi+2 J \int_{r_{\rm min}}^{\infty}\frac{d r}{r^2\sqrt{p_r^2}}\;, \label{chireal}
\end{align}
where the minimum distance $r_\text{min}$ is determined by $p_r(r_{min})=0$, we can obtain the scattering angle up to 4PM order without the radiation-reaction effect, which  is  described as
\begin{eqnarray}
\chi^{\text{Nor}}=\chi^{\text{Nor}}_1\frac{G}{J}+\chi^{\text{Nor}}_2\Big(\frac{G}{J}\Big)^2+\chi^{\text{Nor}}_3 \Big(\frac{G}{J}\Big)^3+\chi^{\text{Nor}}_4 \Big(\frac{G}{J}\Big)^4  \;,
\end{eqnarray}
with
 \begin{eqnarray}
&& \chi_1^{\text{Nor}}=2 m_1 m_2 \frac{1+2\, p_\infty^2}{\sqrt{p_\infty^2}},\nonumber \\ 
&& \chi_2^{\text{Nor}}=\frac{3 \, \pi\, m_1^2 m_2^2}{4} \frac{\left(4+5 \, p_\infty^2\right)  }{\Gamma }, \nonumber \\
&&\chi^{\text{Nor}}_3= 2\, m_1^3\, m_2^3\, \sqrt{p_\infty^2}\, P_{30} +\frac{2}{\pi}\chi^{\text{Nor}}_1 \, \chi^{\text{Nor}}_2-\frac{(\chi^{\text{Nor}}_1)^3}{12} , \nonumber \\
&& \chi^{\text{Nor}}_4=\frac{3 \pi}{4}m_1^4 m_2^4 f_4+\frac{3\pi}{8}\chi^{\text{Nor}}_1 \chi^{\text{Nor}}_3 +\frac{3}{2\pi}(\chi^{\text{Nor}}_2)^2-\frac{3}{4}(\chi^{\text{Nor}}_1)^2 \chi^{\text{Nor}}_2+\frac{\pi}{32}(\chi^{\text{Nor}}_1)^4. \label{4pmchi}\end{eqnarray}
\fi 

It is evident that  the Fourier transform (\ref{FT})  incorporates a factor of $(p_\infty^2)^{-n/2}$, indicating the existence of a singularity at $p_\infty^2 = 0$. Therefore, these results are exclusively applicable to scattering states ($p_\infty^2 > 0$) and cannot be directly utilized for bound states ($p_\infty^2 < 0$). Fortunately,  the challenges encountered in mathematics align precisely with those investigated by Hawking in his research on black hole radiation \cite{Hawking75}.  By employing Hawking's method \cite{Hawking75}, which extends positive frequency solutions to encompass negative frequency solutions, we can analytically continue the results applicable to  $p_\infty^2 > 0$ 
into the realm of $p_\infty^2 < 0$. This is accomplished by substituting $p_\infty^2$ with $p_\infty^2 e^{-i \pi}$. Such a modification enables the effective application of the metric obtained from scattering angles to the analysis of binary black hole mergers. Notably, by substituting $p_\infty^2$ with $|p_\infty^2|$ in the formulation, we find that the effective metric can be suitably applied to both scattering and bound states.

Therefore, the effective metric can be expressed as
\begin{eqnarray}
ds_{\text{eff}}^2=g^{\text{eff}}_{\mu\nu}d x^\mu d x^\nu=\frac{\Delta}{r^2} dt^2-\frac{r^2}{\Delta}dr^2- r^2(d\theta^2+\sin^2\theta d\varphi^2),\nonumber \\ \label{Mmetric}
\end{eqnarray}
with
\begin{eqnarray}
&& \Delta=r^2- 2 GM r+\sum_{i=2}^\infty a_i \frac{(GM\big)^i}{r^{i-2}},
\end{eqnarray}
where the coefficients $a_i$ are
\begin{eqnarray} \label{parameters1}
a_2&=&\frac{3 (\Gamma-1)\left(4+5 |p_\infty^2|\right) }{\left(2+3 |p_\infty^2|\right) \Gamma },\nonumber \\
a_3&=& 
\frac{3 }{2 \left(3+4 |p_\infty^2|\right)}\bigg\{\frac{1}{\left(2+3 |p_\infty^2|\right) \Gamma }\Big[108+3
|p_\infty^2| \Big(85+|p_\infty^2| (50\nonumber \\ &-&32 \Gamma )-56 \Gamma \Big)-74 \Gamma \Big]-\frac{2 \,T_3^p}{\sqrt{|p_\infty^2|}}-2 \tilde{P}_{30}
\bigg\},\nonumber \\
a_4&=&  \frac{\Gamma-1
}{4  (4+5 |p_\infty^2|) \Gamma ^3}\bigg\{\frac{560}{|p_\infty^2|}+16 \Big[105+8 \Gamma  (1+\Gamma )\Big]+3 |p_\infty^2| 
\nonumber \\ &\times& 
 \Big[385+43 \Gamma  (1+\Gamma )\Big]\bigg\}
+ \frac{4}{ (4+5 |p_\infty^2|)} \bigg\{\frac{|p_\infty^2| }{48}
 \Big(-390 a_2\nonumber \\ &
  +& 51 a_2^2-164 a_3\Big)+ \Big[(a_2-8) a_2-\frac{10\, a_3}{3}\Big]-\frac{T_4^{\nu}}{|p_\infty^2|}\nonumber \\ & 
  +& \frac{2 \left(1+2 |p_\infty^2|\right) T_3^p}{|p_\infty^2|^{3/2}}-\frac{8\, T_4^p}{3 \pi |p_\infty^2| }\bigg\}, 
     \end{eqnarray}
in which $\tilde{P}_{30}$, $T_3^{p}$, $T_4^{p}$, and $T_4^{\nu}$, for clarity, presented in Appendix \ref{Tp}. It should be pointed out that the metric (\ref{Mmetric}) is of Type-D, allowing us to employ standard general relativistic methods to calculate the radiation reaction force and waveforms.
 
\section{Energy flux, radiation-reaction force and waveform}\label{RRFWF}

The  RRF associated with the ``plus" and ``cross" modes of the GWs emitted by coalescing binaries  is described by   \cite{Buonanno2006}
\begin{eqnarray}  \label{fe11}
\hat{\bm{\mathcal{F}}}\big[g^{\text{eff}}_{\mu\nu}\big]=\frac{1}{\nu  M  \Omega |\vvr\times
   \vP|}\frac{dE}{dt}\vP, 
\end{eqnarray}   
where $\Omega =  |\vvr\times\dot{\vvr}|/r^2$ represents the dimensionless orbital frequency, and $dE/dt$ is the energy flux of the GWs radiated to infinity.  The energy flux can be described by~\cite{Ref:poisson,TagoshiSasaki745}
\begin{eqnarray}\label{de}
\frac{dE}{dt} & = &\lim_{r\rightarrow\infty}\left[\frac{r^2}{4\pi G \omega^2}\int_\theta\int_\varphi\sin\theta\;d\theta\;d\varphi\left| \psi^{B}_4\right|^2\right] .
\end{eqnarray}

On the other hand, by using $\psi^{B}_4=\frac{1}{2}(\ddot h_{+} - i\ddot h_{\times}) $, we can find the waveform \cite{Kidder}
\begin{eqnarray}
h_{+}-i h_{\times}=\sum_{l=2}^{\infty} \sum_{m=-l}^{l} h^{lm}\frac{ \  _{-2}Y^{lm}(\theta, \varphi)}{\sqrt{2\pi}} .\label{waveform}
\end{eqnarray}

The discussions show that, to determine the RRF and waveform, we must identify the null tetrad component of the gravitational perturbed Weyl tensor \( \psi^{B}_4 \) with source terms in the effective spacetime. In this section, we will derive a decoupled, variable-separated Teukolsky-like equation for $\psi^B_4$, construct a formal solution to this equation, and then present the RRF and  waveform.

\subsection{Decoupled and variable-separated equation for \texorpdfstring{ $\psi^B_{4}$}{} in the effective spacetime}

In the effective spacetime (\ref{Mmetric}), introducing a null tetrad defined as 
\begin{eqnarray}\label{NTetrad}
\nonumber &&l^{\mu}=\Big\{\frac{r^{2}}{\Delta}, \, 1 ,\, 0, \, 0\Big\}, \\
\nonumber &&n^{\mu}=\frac{1}{2}\Big\{1,\, -\frac{\Delta}{r^2},\, 0,\, 0\Big\}, \\
\nonumber &&m^{\mu}=\frac{1}{\sqrt{2}\, r }\Big\{0,\, 0,\, 1,\, \frac{{i}}{\sin\theta}\Big\}, \\
 &&\overline{m}^{\mu}=\frac{1}{\sqrt{2}\, r }\Big\{0,\, 0,\, 1,\, -\frac{{i}}{\sin \theta}\Big\},
\end{eqnarray}
we have
\begin{eqnarray}\label{spin}
&&\rho=-\dfrac{1}{r}, \quad\mu=-\dfrac{\Delta_{r}}{2 r^3}, \quad\gamma=\dfrac{\Delta_{r}^{\prime}}{4r^2}+\mu, \nonumber \\  
&&\alpha=-\dfrac{\cot\theta}{2\sqrt{2} r } , \quad\beta=\dfrac{\cot\theta}{2\sqrt{2} r },\nonumber \\ 
&&\Psi_{2}=\frac{1}{12 r^{4 } }\Big(12\Delta-6 r \Delta^{\prime} + r^{2} \Delta^{\prime\prime}-2 r^{2} \Big),\nonumber \\ 
 && \phi_{11}=\frac{1}{8 r^{4} }\Big(4\Delta-4 r \Delta^{\prime} + r^{2} \Delta^{\prime\prime}+2 r^2  \Big), 
 \end{eqnarray}
here and hereafter, $^\prime$ represents a derivative with respect to $r$, and all other spin coefficients, tetrad components of the Weyl tensor, and tetrad components of the tracefree Ricci tensors  are equal to zero.
Then the Newman-Penrose formalism \cite{Carmeli} show us  that there are three equations related to $\psi_{4}$:
\begin{align}
&\Delta\lambda-\overline{\delta}\nu=-(\mu+\overline{\mu})\lambda-(3\gamma-\overline{\gamma})\lambda+(3\alpha+\overline{\beta})\nu-\psi_{4}, \label{Eq1}\\
 \nonumber&\overline{\delta} \psi_{3}-D\psi_{4}+\overline{\delta} \phi_{21}-\Delta \phi_{20}
 =3 \lambda \psi_{2}-2 \alpha \psi_{3}-\rho \psi_{4}-2 \nu \phi_{10}\nonumber \\ & \ \  +2 \lambda \phi_{11} +(2 \gamma-2 \overline{\gamma}+\overline{\mu}) \phi_{20}-2\alpha\phi_{21}-\overline\sigma \phi_{22}, \label{Eq2} \\ &\Delta \psi_{3}-\delta \psi_{4}+\overline{\delta} \phi_{22}-\Delta \phi_{21} =3 \nu \psi_{2}-2(\gamma+2 \mu) \psi_{3}+4 \beta \psi_{4}\nonumber \\ &\ \ -2 \nu \phi_{11}-\overline{\nu} \phi_{20} +2 \lambda \phi_{12}+2(\gamma+\overline{\mu}) \phi_{21}-2( \overline{\beta}+ \alpha) \phi_{22}.\label{Eq3}\end{align}
where $D=l^\mu \partial_\mu$,   ${\bf{\Delta}}=n^\mu \partial_\mu$,  $\delta=m^\mu \partial_\mu$ and  $\bar\delta=\bar{m}^\mu \partial_\mu$.   We \cite{Jing} have shown that, in the effective background spacetime, the gravitational perturbation described by
\begin{align}\label{Pmetric}
 g_{\mu \nu}=g^{\text{eff}}_{\mu\nu}+\varepsilon h_{\mu \nu}^{B}\;,
\end{align}
can be achieved by perturbing all null tetrad quantities, where  $\varepsilon$ is a small quantity. Then, from Eqs. (\ref{Eq1}), (\ref{Eq2}) and (\ref{Eq3}), retaining $\varepsilon$ only to first order, we derive the following perturbation equations:
\begin{align}\label{Eq11}
 &\psi_{4}^{B}+({\bf{\Delta}}+3\gamma-\overline{\gamma}+\mu+\overline{\mu})\lambda^{B}-(\overline{\delta}+3\alpha+\overline{\beta}+\pi-\bar{\tau})\nu^{B}=0,\\
 &\nu^{B} (3\psi_{2}-2\phi_{11})-({\bf{\Delta}}+2\gamma+4\mu)\psi_{3}^{B}+(\delta+4\beta-\tau)\psi_{4}^{B}\nonumber \\&\ \ \ \ +({\bf{\Delta}}+2\gamma+2\overline{\mu})\phi_{21}^{B}-(\overline{\delta}+2\alpha+2 \overline{\beta}-\bar{\tau})\phi_{22}^{B}=0,\label{Eq22}\\
 &\lambda^{B}(3\psi_{2}+2\phi_{11})-(\bar{\delta}+2\alpha+4\pi )\psi_{3}^{B}+(D+4\epsilon-\rho)\psi_{4}^{B}\nonumber \\&\ \ \ \ -(\overline{\delta}+2\alpha-2\bar{\tau})\phi_{21}^{B}+({\bf{\Delta}}+2\gamma -2\overline{\gamma}+\overline{\mu})\phi_{20}^{B}=0,\label{Eq33}
 \end{align}
where all quantities without and with the superscript $B$ represent the background and perturbation quantities, respectively.  

Applying the operator $({\bf{\Delta}} + 3\gamma - \bar{\gamma} + 4\mu + \bar{\mu})$ to Eq. (\ref{Eq33}) and the operator $(\bar{\delta} + 3\alpha + \bar{\beta} - \bar{\tau} + 4\pi)$ to Eq. (\ref{Eq22}), and subsequently subtracting one equation from the other, we derive the perturbational equation in an explicit form:
\begin{align}
&\left[({\bf{\Delta}}+3\gamma-\bar{\gamma}+4\mu+\bar{\mu})(D+4\epsilon-\rho)\right. \notag\\
&-\left.(\bar{\delta}+3\alpha+\bar{\beta}-\bar{\tau}+4\pi)(\delta+4\beta-\tau)-3\Psi_{2} -12 f \Lambda \right]\Psi_{4}^{B} \notag\\
&=T^B_{4} +{\cal{G}}_4^B,
\label{eq06J1}
\end{align}
with
\begin{align}
T^B_{4}&=({\bf{\Delta}}+3\gamma-\bar{\gamma}+4\mu+\bar{\mu})
\Big[(\bar{\delta}-2\bar{\tau}+2\alpha)\Phi_{21}^{B}
-({\bf{\Delta}}+2\gamma\notag\\
& 
-2\bar{\gamma}+\bar{\mu})\Phi_{20}^{B}\Big]
-(\bar{\delta}+3\alpha+\bar{\beta}-\bar{\tau}+4\pi)
\Big[(\bar{\delta}-\bar{\tau}+2\bar{\beta}\notag\\
& 
+2\alpha)\Phi_{22}^{B}
-({\bf{\Delta}}+2\gamma+2\bar{\mu})\Phi_{21}^{B}\Big],\\
 {\cal{G}}_4^B&=\Big[({\bf{\Delta}}+3\gamma-\bar{\gamma}+4\mu+\bar{\mu})(\bar{\delta}+2\alpha+4\pi)
-(\bar{\delta}+3\alpha+\bar{\beta}\nonumber \\ 
& 
-\bar{\tau}+4\pi)({\bf{\Delta}}+2\gamma+4\mu)\Big] \psi_{3}^{B}
-4\Big[({\bf{\Delta}}+3\gamma-\bar{\gamma}+2\mu\nonumber \\ 
&
+2\bar{\mu})(\Phi_{11}\lambda^{B})+\nu^{B}(\bar{\delta}+\pi-\bar{\tau})\Phi_{11}+\frac{1}{2} (\Phi_{11}+6 f \Lambda)\Psi_{4}^{B}\Big],\end{align}
where $\Lambda=-\frac{R}{24}=\frac{2-\Delta''}{24 r^{2}}$ ($R$ is the Ricci scalar curvature), and the factor $f$ should be fixed in the specific physical systems. 

It is important to note that four functions are involved in the three equations (\ref{Eq11}), (\ref{Eq22}), and (\ref{Eq33}): $\nu^{B}$, $\lambda^{B}$, $\Psi_{3}^{B}$, and $\Psi_{4}^{B}$. Therefore, we can impose a gauge condition expressed as 
\begin{eqnarray}
{\cal{G}}_4^B=0. \label{gauge}
\end{eqnarray}
In Appendix \ref{APPC}, we demonstrate that we can always identify a null tetrad that satisfies the specified gauge condition by performing a Class I rotation \cite{Chandrasekhar}.
Consequently, from Eq. (\ref{eq06J1}), we establish that the decoupled equation for $\Psi_{4}^{B}$ can be expressed as
\begin{align}
&\Big[({\bf{\Delta}}+3\gamma-\bar{\gamma}+4\mu+\bar{\mu})(D+4\epsilon-\rho) -(\bar{\delta}+3\alpha+\bar{\beta}-\bar{\tau}\notag\\
&
+4\pi)(\delta+4\beta-\tau)-3\Psi_{2} -12 f \Lambda \Big]\Psi_{4}^{B}=T^B_{4}.
\label{decouple4}
\end{align}

By taking $\Psi_{4}^{B}=r^{-4}  \phi_{4}^{B}$, we can express the decoupled equation for $\psi_{4}^{B}$, as described in Eq. (\ref{decouple4}), in the form of a Teukolsky-like equation given by
 \begin{align}
&\frac{r^{4}}{ \Delta }\frac{\partial^{2} \phi_{4}^{B}}{\partial t^{2}}
+\Big(\frac{2\, r^{2} \Delta' }{ \Delta }-8r \Big)\frac{\partial \phi_{4}^{B}}{\partial t}
-\Delta^2  \frac{\partial}{\partial r}\Big(\frac{1}{ \Delta }\frac{\partial \phi_{4}^{B}}{\partial r}\Big)
\notag\\
&{}
-\frac{1}{\sin\theta}\frac{\partial}{\partial\theta}\Big(\sin\theta\frac{\partial \phi_{4}^{B}}{\partial\theta}\Big)
-\frac{1}{\sin^{2}\theta}\frac{\partial^{2} \phi_{4}^{B}}{\partial\varphi^{2}}
+\frac{4i\cos\theta}{\sin^2\theta}\frac{\partial \phi_{4}^{B}}{\partial\varphi}
\notag\\
&{}
+\Big[4\cot^{2}\theta+2+\Big(\frac{1}{2}-f\Big)(2-\Delta'')\Big] \phi_{4}^{B} \equiv  {\cal T}^{(-2)},
\label{PDE}
\end{align} 
with
\begin{align}
\nonumber {\cal T}^{(-2)}&=-\frac{4\pi G}{r^{3}}\bigg\{\mathscr{L}_{-1}\Big[r^{4} \mathscr{L}_{0} \big(r^{3}T_{nn}\big)  \Big]+\frac{\Delta ^{2}}{2}\mathscr{D}_{0}^{\dagger}\Big[r^{4}\mathscr{D}_{0}^{\dagger}\Big(
r T_{\bar{m}  \bar{m}}\Big)  \Big]\\   &+ \frac{\Delta ^2}{\sqrt{2}}\Big\{\mathscr{D}_{0}^{\dagger}\Big[
\frac{r^{2}}{\Delta }
\mathscr{L}_{-1}\Big(
r^{4}T_{\bar{m} n}\Big) \Big]+\mathscr{L}_{-1}\Big[
r^{2}
\mathscr{D}_{0}^{\dagger}\Big(
\frac{r^{4}}{\Delta }T_{ \bar{m} n}\Big)  \Big]\Big\}\bigg\},
\label{T4TT} 
\end{align}
where 
 \begin{align}
\nonumber &\mathscr{D}_{n}=\partial_{r}-\frac{i K}{\Delta_{r}}+n\frac{\Delta'_{r}}{\Delta_{r}}, \qquad\mathscr{D}_{n}^{\dagger}=\partial_{r}+\frac{iK}{\Delta_{r}}+n\frac{\Delta'_{r}}{\Delta_{r}}, \\
 &\mathscr{L}_{n}=\partial_{\theta}-\mathscr{Q}+n\cot\theta, \qquad\mathscr{L}_{n}^{\dagger}=\partial_{\theta}+\mathscr{Q}+n\cot\theta,
\end{align}
in which $
 K=r^2 \omega ,$ $\mathscr{Q}=-\frac{m}{\sin\theta} .$

In the non-homogeneous case, we can express \( \phi_{4}^{B} \) and \( T^{(-2)} \) as expansions in terms of the functions \( {}_{s}Y^{a\omega}_{lm}(\theta) \)
\begin{align}
&\phi_{4}^{B}=\int d\omega \sum_{l,m} R^{(-2)}_{lm\omega}(r){}^{}_{s}Y^{a\omega}_{lm}(\theta)e^{-i\omega t}e^{im\varphi},
\\
&{\cal T}^{(-2)}=\int d\omega \sum_{l,m} {\cal T}^{(-2)}_{lm\omega}(r){}^{}_{s}Y^{a\omega}_{lm}(\theta)e^{-i\omega t}e^{im\varphi}.\label{TTT}
\end{align}
Subsequently, the separated equations for Eq.   (\ref{PDE}) can be expressed as 
\begin{align}
&\Delta^{2}\frac{d}{dr}\bigg(\frac{1}{\Delta}\frac{d R^{(-2)}_{lm\omega}}{dr}\bigg)
+\bigg[\frac{K^{2}+2 i  K\Delta'}{\Delta}+\Big(f-\frac{1}{2}\Big) (2-\Delta'') \notag\\
&{}
-8 i  \omega r-\boldsymbol{\lambda}\bigg]R^{(-2)}_{lm\omega}={\cal T}^{(-2)}_{lm\omega},
\label{defTE1}
\\
&\frac{1}{\sin\theta}\frac{d}{d\theta}\bigg(\sin\theta\frac{d {}_{-2}Y^{a\omega}_{lm}}{d\theta}\bigg)
-\bigg(\frac{m^{2}-4  m\cos\theta}{\sin^{2}\theta} +4\cot^2\theta\notag\\
&{}
+2-\boldsymbol{\lambda}\bigg) {}_{-2}Y^{a\omega}_{lm}=0,
\label{swsh}
\end{align}
with
\begin{align}
&{\cal T}^{(-2)}_{\ell m \omega}(r)=\frac{1 }{2\pi } \int_{-\infty }^{+\infty} dt \int d\Omega \ {\cal T}^{(-2)} \ e^{i(\omega t-m \varphi)} \frac{ \ _{-2}Y^*_{\ell m}(\theta ) }{\sqrt{2\pi} }. 
\label{VrgenTT1}
\end{align}
where  $\boldsymbol{\lambda}=(l+2)(l-1)$. 
Eq. (\ref{defTE1}) simplifies to the radial Teukolsky equation in Schwarzschild spacetime for the special case in which $a_2=a_3=a_4=0$, resulting in $\Delta=r^{2}-2Mr$.

\subsection{ The source  of the gravitational radiation}

We introduce the null tetrad symbol \( Z_{a\mu}=(l_\mu,\, n_\mu,\, m_\mu,\, \bar{m}_\mu) \) for the effective background spacetime. In this context, \( l_\mu, n_\mu, m_\mu, \) and \( \bar{m}_\mu \) are defined by Eq. (\ref{NTetrad}). The projections of a tensor \( A^{\mu\nu...} \) onto the null tetrad can be represented as \( A_{ab...}=A^{\mu\nu...}Z_{a\mu}Z_{b\nu}... \). Furthermore, we define the spin coefficients as \( \gamma_{abc}=\nabla_\nu Z_{a\mu}Z_{b}^{\mu}Z_{c}^{\nu} \) \cite{Carmeli,Chandrasekhar}. It follows that the projections of the energy-momentum tensor onto the null tetrad are given by
\begin{align}
T_{a b}&=Z_{a\mu}Z_{b\nu}T^{\mu\nu}(x)=\int d\tau\bigg[\frac{ m_0 v_{(a}v_{b)}}{\sqrt{-g}}\delta^{(4)}(x-z(\tau))\bigg],
\label{energy-momentum tensor Null}
\end{align}
where $v^{\mu}(\tau)=d z^{\mu}(\tau)/d\tau$. In the context of circular equatorial orbits  for a particle in the effective  background  spacetime, Eq. (\ref{energy-momentum tensor Null}) shows that the tetrad components of the energy-momentum tensor employed in \eqref{T4TT} can be expressed as
\begin{eqnarray}
&&T_{n n}=   \frac{C_{n n} }{ \sin\theta}
 \delta(\theta-\theta(t)) \delta(\varphi-\varphi(t)),
\nonumber\\
&&T_{{\bar m} {\bar m}}= \frac{C_{{\bar m} {\bar m}} }{ \sin\theta}
 \delta(\theta-\theta(t)) \delta(\varphi-
\varphi(t)),\nonumber\\
&&T_{n {\bar m} }=   \frac{ C_{{n \bar m} } }{ \sin\theta}
 \delta(\theta-\theta(t)) \delta(\varphi-\varphi(t)),
 \label{tij}
\end{eqnarray}
with
\begin{align}
&C_{n n}=B_{n n}\delta(r-r(t))=\frac{m_0 v_n v_n}{ r^2 \dot{t} }\delta(r-r(t)),\nonumber \\ 
&C_{n \bar m}=B_{n \bar m}\delta(r-r(t))=\frac{1}{ 2 r^2 \dot{t} }\Big( m_0 (v_{n}v_{\bar m}+v_{\bar m}v_{n})\Big)\delta(r-r(t)),\nonumber \\
&C_{\bar m\bar m}=B_{\bar m\bar m}\delta(r-r(t))=\frac{ m_0 v_{\bar m}v_{\bar m}}{ r^2 \dot{t} }\delta(r-r(t)), \end{align}
where $ v_n=n_\mu v^{\mu},$ $ v_{ m}=m_\mu v^{\mu},$  $   v_{\bar m}=\bar{m}_\mu v^{\mu}$, and $\dot t=dt/d\tau$. 

Utilizing Eqs.~(\ref{T4TT}), (\ref{TTT}), and (\ref{tij}), and applying integration by parts, we derive the following expression
\begin{eqnarray}
{\cal T}^{(-2)}_{\ell m \omega}
&=&
\frac{4m_0 G}{\sqrt{2\pi}}\int^{\infty}_{-\infty}
dt\int d\theta e^{i\omega t-im\varphi(t)} \nonumber\\
&
\times&\bigg\{-\frac{1}{ 2}\mathscr{L}_1^{\dag} \Big[ r^{4} \mathscr{L}_2^{\dag}\Big( \frac{\ _{-2}Y_{\ell m}}{r^{3}}\Big)
\Big]
C_{n n}r^{3} 
\delta(\theta-\theta(t)) \nonumber\\
&
+&\frac{\Delta^2 }{  \sqrt{2} r}
\Big[ \mathscr{L}_2^{\dag} \big( _{-2}Y_{\ell m}\big)\Big]\mathscr{D}_{0}^{\dagger} \Big[
\frac{C_{{\overline m} n} r^{4} }{\Delta}
\delta(\theta-\theta(t)) \Big] \nonumber\\
&
+&\frac{1}{ 2\sqrt{2} }
\mathscr{L}_2^{\dag}\Big[\frac{ \ _{-2}S_{\ell m}}{ r^{3}}\frac{\partial }{\partial r}\Big(r^2\Big) \Big]
C_{{\overline m} n}\Delta \ r^{4}
\delta(\theta-\theta(t)) \nonumber\\
&
-&\frac{\Delta^2 }{ 4 r^{3}} \big( _{-2}Y_{\ell m} \big) \mathscr{D}_{0}^{\dagger}\Big[r^{4}
\mathscr{D}_{0}^{\dagger} \Big(r C_{{\overline m}{\overline m}}
\delta(\theta-\theta(t))\Big) \Big]
\bigg\}. 
\label{Tgen}
\end{eqnarray}
 For a source constrained within a finite range of $r$, Eq.~(\ref{Tgen}) can be reformulated as
\begin{align}
{\cal T}^{(-2)}_{\ell m \omega}&=m_0 G \int^{\infty}_{-\infty}dt 
e^{i\omega t-i m \varphi(t)}
\Delta  ^2\Big\{\big(A_{nn\,0} B_{nn}+A_{{\bar m}n\,0} B_{{\bar m}n}\notag\\
&{}
+
A_{{\bar m}{\bar m}\,0} B_{{\bar m}{\bar m}}\big) \delta(r-r(t)) +\big[\big(A_{{\bar m}n\,1} B_{{\bar m}n}+A_{{\bar m}{\bar m}\,1} B_{{\bar m}{\bar m}} \big)
\notag\\
&{}
\times\delta(r-r(t))\big]
' +\big(A_{{\bar m}{\bar m}\,2}B_{{\bar m}{\bar m}}\delta(r-r(t))\big)
''\Big\} ,
\label{NTgenTTsl}
\end{align}
where the coefficients $A_{ij\,a}$ can be found in Appendix \ref{AijA}. 

\subsection{ Formal solution of Teukolsky-like equation for   \texorpdfstring{ $\psi^B_{4}$}{}  }

In this subsection, we will solve the Teukolsky-like equation, Eq. (\ref{defTE1}),  utilizing the Green function method. The asymptotical homogeneous solutions of Eq.~\eqref{defTE1} is \begin{eqnarray}
R^{\rm in (-2)}_{asy}
\to \left\{\begin{array}{cc}
B^{\rm trans}_{\ell m\omega}\Delta^2 e^{-i \omega r^*},
\ \ \ \ \ \ \ \ \ \ \ \ \ \ \ \    & \ \ \text{for} \ \ r\to r_+, \ \ \ \\
r^3 B^{\rm ref}_{\ell m\omega}e^{i\omega r^*}+
\frac{1}{r}B^{\rm inc}_{\ell m\omega} e^{-i\omega r^*},
 & \text{for} \  r\to +\infty,
\end{array}\right.
\label{Kk}
\end{eqnarray}

\begin{eqnarray}
R^{\rm up (-2)}_{asy}
\to \left\{\begin{array}{cc}
C^{\rm up}_{\ell m\omega} e^{i \omega r^*}+
\Delta^2 C^{\rm ref}_{\ell m\omega} e^{-i \omega r^*},
& \ \ \ \text{for} \ \ r\to r_+, \ \ \ \ \\
C^{\rm trans}_{\ell m\omega} r^3 e^{i\omega r^*},
\ \ \ \ \ \ \ \ \ \ \ \ \ \ \ \  & \text{for} \  r\to +\infty,
\end{array}\right.
\label{Kkk}
\end{eqnarray}
where $r^*$ denotes the tortoise coordinate defined by $ r=\int \frac{r^2}{ \Delta}dr $.
Subsequently, the inhomogeneous solution of the radial equation~\eqref{defTE1} can be constructed as
\begin{align}
\label{EqTS0S}
R^{ (-2)}_{\ell m\omega}(r)&=\frac{1}{2 i \omega C^{\rm trans}_{\ell m\omega}
     B^{\rm inc}_{\ell m\omega}}
 \bigg[R^{\rm up  (-2)}_{\ell m\omega}(r)\int^r_{r_+}d\tilde{r}
\frac{ R^{\rm in  (-2)}_{\ell m\omega}(\tilde{r} )
{\cal T}^{(-2)}_{\ell m\omega}(\tilde{r} ) }{ \Delta^{2} (\tilde{r} ) }\nonumber \\ &+ R^{\rm in (-2)}_{\ell m\omega} (r) \int^\infty_{r}d\tilde{r}
\frac{R^{\rm up  (-2)}_{\ell m\omega}(\tilde{r} ) {\cal T}^{ (-2)}_{\ell m\omega}(\tilde{r} ) }{ \Delta^{2}(\tilde{r} ) }\bigg],
\end{align}
where $R^{\rm up  (-2)}_{\ell m\omega}(\tilde{r} ) $ and $R^{\rm in  (-2)}_{\ell m\omega}(\tilde{r} ) $ represent the homogeneous solutions of the radial equation \eqref{defTE1}. 
Consequently, the inhomogeneous solution for Eq. \eqref{defTE1}  at the infinity can be expressed as
\begin{align}
R^{(-2)}_{\ell m\omega}(r\to\infty) & \to \frac{r^3e^{i\omega r^*} }{ 2i\omega
   B^{\rm inc}_{\ell m\omega}}
\int^{\infty}_{r_+}d\tilde{r} \frac{
R^{\rm in (-2)}_{\ell m\omega}(\tilde{r} ){\cal T}^{(-2)}_{\ell m\omega}(\tilde{r} )}{  \Delta^{2}}
\nonumber \\ & \equiv \tilde Z^{I (-2)}_{\ell m\omega} r^3 e^{i\omega r*}\,.
\label{Infinfty}
\end{align}
By substituting Eq.~(\ref{NTgenTTsl}) into Eq.~(\ref{Infinfty}) and subsequently performing integration by parts, we obtain
\begin{align}
\tilde Z^{ (-2)}  _{\ell m\omega}&=
\frac{m_0 G}{2i\omega B^{\rm inc}_{\ell m\omega}}
\int^{\infty}_{-\infty}dt e^{i\omega t-i m \varphi(t)}
\Big[ A^{(-2)}_0
R^{\rm in(-2)}_{\ell m\omega}\nonumber \\ & - A^{(-2)}_1
\Big(R^{\rm in(-2)}_{\ell m\omega}\Big)' + A^{(-2)}_2 \Big(R^{\rm in(-2)}_{\ell m\omega}\Big)'' \Big], 
\label{ZZSch}
\end{align}
where $A^{(-2)}_0 =\big(A_{nn\,0} B_{nn}+A_{{\bar m}n\,0} B_{{\bar m}n}+
A_{{\bar m}{\bar m}\,0} B_{{\bar m}{\bar m}}\big),$ $
 A^{(-2)}_1 =\big(A_{{\bar m}n\,1} B_{{\bar m}n}+A_{{\bar m}{\bar m}\,1} B_{{\bar m}{\bar m}} \big),$ $
 A^{(-2)}_2 =A _{{\overline m}{\overline m}\,2}B_{{\overline m}{\overline m}}$.
 
 When considering only circular orbits, \( r(t) \) in Eq. (\ref{ZZSch}) is independent of time, allowing us to set \( r(t) = r_0 \). Along the geodesic trajectory, we also have \( \theta(t) = \theta_0 \) and \( \varphi(t) = \Omega\, t \), where \( \Omega \) represents the angular velocity. By proceeding with the integration for Eq. (\ref{ZZSch}), and utilizing Eqs.~(\ref{TTT}) and (\ref{Infinfty}), we derive the formal solution for \( \psi_4^B \) at infinity as follows
\begin{align}
\psi^{B}_4&=\frac{1}{ r}\sum_{\ell m n}
\frac{\pi m_0 G}{i\omega_n B^{\rm inc}_{\ell m\omega_n}}
\bigg[ A^{(-2)}_0
R^{\rm in(-2)}_{\ell m\omega}- A^{(-2)}_1
\Big(R^{\rm in(-2)}_{\ell m\omega}\Big)' \nonumber \\ & + A^{(-2)}_2 \Big(R^{\rm in(-2)}_{\ell m\omega}\Big)'' 
\bigg]_{r_0,\theta_0} \frac{{}_{-2}Y_{\ell m} }{ \sqrt{2\pi}}
e^{i\omega_n(r^*-t)},  \ \  \text{for}\ \  (r \to \infty). \label{psi411}
\end {align}

\subsection{Radiation-reaction force and waveform}\label{RRFWFB}

By utilizing Eqs. (\ref{fe11}), (\ref{de}) and (\ref{psi411}), for the quasicircular cases without precession, noting that $|\vvr\times \vp|\approx p_\varphi $, we find that the reduced RRF that appeared in the Hamiltonian equation (\ref{HEq}) can be expressed explicitly as
\begin{align}
\hat{\bm{\mathcal{F}}}[g^{\text{eff}}_{\mu\nu}]&=\frac{1}{\nu M_0 \Omega}\sum_{\scriptstyle \ell=2}^{\infty}\sum_{m=1}^\ell
\dfrac{2 \pi m_0^2 }{ G\omega_n^4}\Bigg| \frac{G}{B^{\rm inc}_{\ell m\omega_n}}
\bigg[ A^{(-2)}_0
R^{\rm in(-2)}_{\ell m\omega_n}(r)\nonumber \\ &- A^{(-2)}_1 \Big(R^{\rm in(-2)}_{\ell m\omega_n}(r)\Big)' + A^{(-2)}_2\Big(R^{\rm in(-2)}_{\ell m\omega_n}(r)\Big)'' \bigg]_{r_0,\theta_0}
 \Bigg|^2\frac{\vp}{p_{\varphi}},
 \label{FFdE1}
\end{align}
indicating that the reduced RRF is constructed in terms of the effective spacetime.

From Eqs. (\ref{waveform}) and (\ref{psi411}),  we find the waveform
\begin{eqnarray}
h^{l m}&=&\frac{1}{ r }
\frac{2 \pi m_0 G}{i\omega_n^3 B^{\rm inc}_{\ell m\omega_n}}
\bigg[ A^{(-2)}_0
R^{\rm in(-2)}_{\ell m\omega_n}- A^{(-2)}_1
\Big(R^{\rm in(-2)}_{\ell m\omega_n}\Big)' \nonumber \\ &+& A^{(-2)}_2 \Big(R^{\rm in(-2)}_{\ell m\omega_n}\Big)'' 
\bigg]_{r_0,\theta_0} e^{i\omega_n(r^*-t)}.
\label{hform}
\end{eqnarray}

The preceding discussion indicates that all formulas and quantities, including the effective Hamiltonian (\ref{HHH}), the energy flux (\ref{de}), the reduced RRF (\ref{FFdE1}), and the waveform (\ref{hform}), are derived from the effective metric. Consequently, the EOB theory for spinless binaries can be regarded as a self-consistent theoretical framework.

\section{Compared the results with NR data }

It is widely recognized that GW events produced by coalescing binary systems can be accurately characterized by waveforms generated through NR. Not only does NR provide precise waveform templates for GW signals, but it also offers vital calibration support for other approximation methods. 

The analysis of the energetics can done via the gauge-invariant  relation between the binding energy $E_b$ and  the total angular momentum $j$, which are computed as \cite{PhysRevD.93.044046,PhysRevLett.108.131101,PhysRevD.98.104057}
\begin{eqnarray}
E_b      & \equiv& \dfrac{M_{\rm ADM}^0-\Delta {\cal E}_{\rm rad} -M}{\mu}, \\
j        & \equiv&  \dfrac{{\cal J}^0_{\rm ADM}-\Delta{\cal J}_{\rm rad}}{M\mu},
\end{eqnarray}
where  $(M_{\rm ADM}^0,{\cal J}^0_{\rm ADM})$ denote the total, initial ADM mass-energy and angular momentum of the system, $(\Delta {\cal E}_{\rm rad}, \Delta{\cal J}_{\rm rad})$ denote the energy and angular momentum radiated in GWs, while $M=m_1+m_2$ and $\mu=m_1m_2/M$, where $m_1$ and $m_2$ are the NR measured  initial Christodoulou masses.
We now present a comparison between the binding energy $E_{b}(j)$ calculated using EOB theory and the results from NR simulations since the binding energy is a critical ingredient in the computation of gravitational waveforms \cite{PhysRevD.93.044046,PhysRevLett.108.131101,PhysRevD.98.104057}.

\begin{figure}[htbp]
\includegraphics[height=19.cm,width=22.5cm]{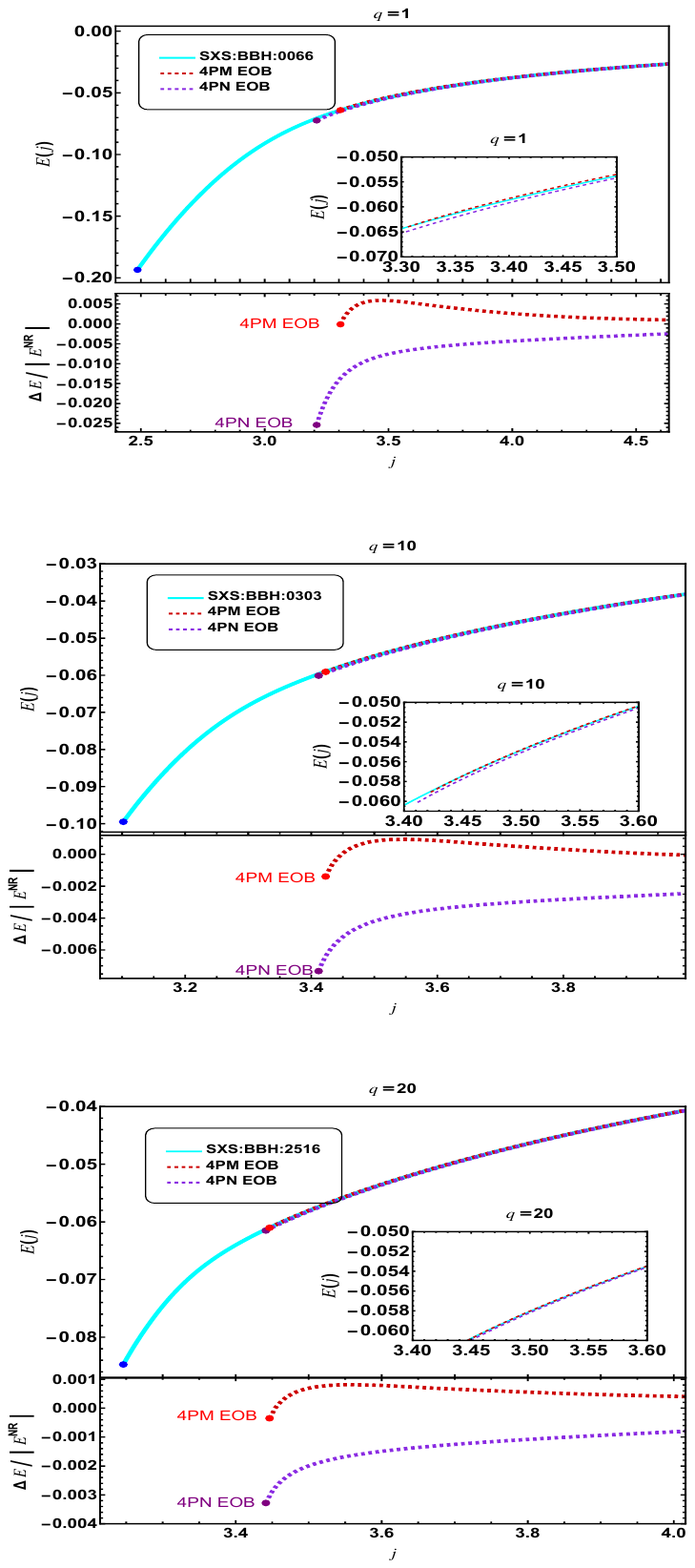}\\
	\caption{ Plots of $E_{b}(j)$ with the mass ratio of $q=1,\  10\ \text{and}\  20$ for  NR data (SXS), 4PM EOB and 4PN EOB,  where pink/purple point denotes the innermost stable circular orbit calculated by 4PM/4PN EOB, and blue point represents the final state of the black hole from NR simulations.}
		\label{fig:sxsbbh006omegaandEj}
	\end{figure}

The topmost plot in the Fig. \ref{fig:sxsbbh006omegaandEj} presents a comparison of the $E_{b}(j) $ curves calculated using EOB theory with NR simulation results for a non-spinning binary black hole system with a mass ratio of $q=1$. The NR simulation data is sourced from the SXS database, specifically the signal labeled \texttt{SXS:BBH:0066}. Since the EOB calculation employs the adiabatic approximation, we only display the results evolved to the innermost stable circular orbit (ISCO). It is evident that the agreement between our calculated curve and the NR results remains within the range of $0 \sim 5$\textperthousand.

Additionally, we compare our calculations with NR simulation data (\texttt{SXS:BBH:0303}) for a binary black hole system with a larger mass ratio,  $q=10$, as shown  in the middle plot of Fig. \ref{fig:sxsbbh006omegaandEj}. The error curves indicates that the difference between our $E_b(j) $ curves and the NR data is less than $1$\textperthousand.
The NR simulation data depicted the bottom plot of Fig. \ref{fig:sxsbbh006omegaandEj} corresponds to the simulation with ID \texttt{SXS:BBH:2156} (with $q=20$), and the error curves in Fig. \ref{fig:sxsbbh006omegaandEj} show that the error is less than $0.8$\textperthousand.

It is apparent that in a binary black hole system, a larger mass ratio corresponds to a higher consistency between the adiabatic gauge-invariant quantity calculated using the EOB method and the NR simulation results. 

We also compared our results with those for the 4PN EOB \cite{PhysRevD.91.084024,khalil2023theoreticalgroundworksupportingprecessingspin}. From this comparison, it is evident that our results exhibit a significantly higher level of precision, representing an improvement of half an order of magnitude over the 4PN EOB theory, as demonstrated in the figure.

\section{ Conclusions and discussions}

The EOB theory represents a pioneering approach to the analysis of the two-body dynamics of compact objects. In the context of spinless black-hole binaries, this theory offers a conceptual framework to map the conservative dynamics of two compact objects with masses (\( m_1 \), \( m_2 \)) onto the dynamics of an effective particle with mass \( m_0 \), orbiting around a massive black hole characterized by mass \( M \). The dynamical evolution within the framework of EOB theory for this system is articulated through Hamilton equations. For the theory to maintain its self-consistency, all formulas and quantities stated in the Hamiltonian equations must derive from the unified foundational physical model. 

We began by deriving the effective metric (\ref{Mmetric}) for spinless binaries up to the fourth PM order. In our calculations, we analytically continued the results applicable to  $p_\infty^2 > 0$ 
into the domain of  $p_\infty^2 < 0$  by employing Hawking's method \cite{Hawking75}. This allowed us to effectively utilize the metric obtained from scattering angles \cite{Jing3} in the analysis of binary black hole mergers.
 
After that, as a compact object binary system generates GWs, we note that the energy of the binary system is lost, leading to the RRF. The reduced RRF is associated with the energy flux of the gravitational radiation, given by \( \frac{dE}{dt} = \frac{1}{16\pi G} \int (\dot{h}_{+}^2 + \dot{h}_{\times}^2) r^2 d \Omega \), as expressed by 
$ \hat{\bm{\mathcal{F}}}=\frac{1}{\nu M_0 \Omega |\vvr\times
   \vP|}\frac{dE}{dt}\vP $ \cite{Buonanno2006}. 
Therefore, to determine the RRF of the ``plus" and ``cross" modes of the GWs produced by the  binaries, we must identify the null tetrad component of the gravitational perturbed Weyl tensor \( \psi^B_{4} = \frac{1}{2}(\ddot h_{+} - i\ddot h_{\times}) \) with source terms in the effective spacetime. Due to the complexity of finding the RRF, we have divided the task into four steps: 1) We first established that, in the effective spacetime, the decoupled equation for \( \psi_{4}^{B} \) described by Eq. (\ref{PDE}), and separated the variables in the Teukolsky-like equation, where the radial equation is given by Eq. (\ref{defTE1}); 2) We identified the sources in the Teukolsky-like equation for \( \psi_4^B \), which relate to Eq. (\ref{energy-momentum tensor Null}); 3) We derived the formal solution of the Teukolsky-like equation presented in Eq. (\ref{psi411}); and 4) We present  the RRF shown by Eq. (\ref{FFdE1}), and the waveform described by Eq. (\ref{hform}) for the ``plus" and ``cross" modes of GWs generated by the spinless binaries.
It is noteworthy that the dynamics of the evolution of spinless binaries within EOB theory represents a self-consistent framework, as all formulae and quantities in the equations of motion are derived from the unified physical model.

The binding energy \cite{PhysRevD.93.044046,PhysRevLett.108.131101,PhysRevD.98.104057} is a critical ingredient in the computation of gravitational waveforms. We performed numerical calculations based on the adiabatic approximation using the EOB theory and compared the results with NR data (SXS). The results show that up to ISCO for $q=1$, the binding energy-angular momentum relation differs from NR results by less than $5$\textperthousand; for $q=10$, the difference is further reduced to within $1$\textperthousand; and for $q=20$, the difference is less than $0.8$\textperthousand. These findings confirm that a larger mass ratio leads to a better agreement between theoretical and NR results. Although the current results are based solely on the adiabatic approximation, they already exhibit a very high consistency with NR. In the future, if we further incorporate the orbital evolution due to radiation reaction, the matching accuracy is expected to be further improved.

It is well-known that a general gravitational waveform template must be constructed for spinning binaries. Our preliminary studies indicate that all necessary conditions outlined in this paper are satisfied for spinning black holes. Subsequently, we will extend this theoretical framework from a non-spinning system to a binary system of spinning black holes. 

\vspace{0.3cm}
\acknowledgments
{ We would like to thank professors S. Chen,  Q. Pan and X. He for useful discussions on the manuscript. This work was supported by the Grant of NSFC Nos. 12035005 and 12475051, and National Key Research and Development  Program of China No. 2020YFC2201400.}  

\newpage
\appendix
\section*{Appendix}

\section{ The coefficients $P_n$ in the Fourier transform of the scattering amplitude (\ref{FT})}\label{Pn}
\begin{eqnarray}
P_0 &=& \left(\frac{\mu}{\Gamma}\right)^2 p_\infty^2, \nonumber \\
P_1 &=& 2M\mu^2\left(\frac{2p_\infty^2 + 1}{\Gamma}\right), \nonumber \\
P_2 &=& \frac{3M^2\mu^2}{2} \left(\frac{5p_\infty^2 + 4}{\Gamma}\right), \nonumber 
 \\
P_3&=&  \frac{M^3 \mu^2 }{\Gamma }    \Bigg\{\frac{17}{2}+9 p_\infty ^2+\frac{\nu}{6}   \Bigg[\frac{18 \Gamma  \left(2
   p_\infty ^2+1\right) \left(5 p_\infty ^2+4\right)}{(\Gamma +1)
   \left(\sqrt{p_\infty ^2+1}+1\right)}-4
   \left(p_\infty ^2+1\right)^{3/2}-108 \left(p_\infty ^2+1\right)\nonumber \\ 
  &-&206
   \sqrt{p_\infty ^2+1}+6\Bigg]+\frac{8 \nu  \left(11+4   p_\infty ^2-4 p_\infty ^4\right) \arcsinh\sqrt{\frac{\sqrt{p_\infty ^2+1}-1}{2}}}{
   \sqrt{p_\infty ^2}}\Bigg\},\nonumber \\ 
P_4&=& -\frac{9\, M^2 \mu^2\,\left(4+5\, p_\infty ^2 \right)^2}{8\, p_\infty ^2}-\frac{2\, M\, \Gamma \,\left(1+2\, p_\infty ^2 \right) P_3}{ p_\infty ^2 }+\frac{4 \, M^2\, \mu^2\, \Gamma^2\, T_{4p}}{3\,\pi \, p_\infty ^2 },  \label{P01234}
\end{eqnarray} 
with
\begin{eqnarray} 
T_{4p} &=&\frac{h[61]}{16 p_\infty ^2  \Gamma ^3}+\frac{ p_\infty^4 \nu }{144
\Gamma ^3} \Bigg\{\frac{36 }{ p_\infty^4}\text{ E}\Big(\frac{\sqrt{1+ p_\infty^2 }-1}{1+\sqrt{1+ p_\infty^2 }}\Big)
\text{ K}\Big(\frac{\sqrt{1+ p_\infty^2 }-1}{1+\sqrt{1+ p_\infty^2 }}\Big) h_{4}-\frac{24 \,\pi^2\, h_{5}}{ p_\infty^2 }\nonumber \\ &-&\frac{126 \text{ E}\Big(\frac{\sqrt{1+ p_\infty^2 }-1}{1+\sqrt{1+ p_\infty^2 }}\Big)^2 h_{2}}{ p_\infty^2  (\sqrt{1+ p_\infty^2 }-1)} 
-\frac{36}{ p_\infty^4}
\text{ K}\Big(\frac{\sqrt{1+ p_\infty^2 }-1}{1+\sqrt{1+ p_\infty^2 }}\Big)^2 h_{3}+\frac{12\, h_{24} \, \text{arccosh}\sqrt{1+ p_\infty^2 }
}{ p_\infty^7}\nonumber \\ 
&+&\frac{18 \,h_{26}\,\text{arccosh}(\sqrt{1+ p_\infty^2 })^2 }{ p_\infty^8}-\frac{h_{62}}{ p_\infty^6 (1+ p_\infty^2 )^{7/2}}-\frac{48\,
h_{23}\, \text{log}(1+ p_\infty^2 )}{ p_\infty^4}\nonumber \\ 
&-&\Big(\frac{12\, h_{6} }{ p_\infty^4}+\frac{36\, h_{16}\,\text{arccosh}(\sqrt{1+ p_\infty^2 })
 }{ p_\infty^5}\Big)\text{log}\Big(\frac{1}{2} \Big(\sqrt{1+ p_\infty^2 }-1\Big)\Big)\nonumber \\ 
&-&\Big(\frac{12\, h_{22} }{ p_\infty^4}+\frac{36\, h_{28} \, \text{arccosh}\Big(\sqrt{1+ p_\infty^2 }\Big)}{ p_\infty^5}\Big)\text{log}\Big(\frac{1}{2} \Big(1+\sqrt{1+ p_\infty^2 }\Big)\Big)\nonumber \\ 
&+&\frac{72 h_{15}
\text{log}(\frac{1}{2} (\sqrt{1+ p_\infty^2 }-1)) \text{log}(\frac{1}{2} (1+\sqrt{1+ p_\infty^2 }))}{ p_\infty^2 }-\frac{48 h_{29} }{ p_\infty^2 } \text{Li}_2(\frac{1-\sqrt{1+ p_\infty^2 }}{1+\sqrt{1+ p_\infty^2 }})\nonumber \\
&-&\frac{288\, h_{27}\,
\text{log}\Big(\frac{1}{2} \Big(1+\sqrt{1+ p_\infty^2 }\Big)\Big)^2}{ p_\infty^2 }-\frac{576\, h_{7} \, \sqrt{ p_\infty^2 } \, \text{Li}_2\Big(\sqrt{\frac{\sqrt{1+ p_\infty^2 }-1}{1+\sqrt{1+ p_\infty^2 }}}\Big)}{ p_\infty^4 (1+\sqrt{1+ p_\infty^2 })}
\nonumber \\
&+&\frac{72\, (2\, \sqrt{ p_\infty^2 }\,
h_{7}- p_\infty^2  (1+\sqrt{1+ p_\infty^2 })\, h_{30})  \text{Li}_2\Big(\frac{\sqrt{1+ p_\infty^2 }-1}{1+\sqrt{1+ p_\infty^2 }}\Big)}{ p_\infty^4 (1+\sqrt{1+ p_\infty^2 })}\Bigg\},
\end{eqnarray} 
where 
$
\text{Li}_2(z) \equiv \int_z^0 dt \, \frac{\log(1-t)}{t}\,, $ $ \mathrm{K}(z) \equiv 
 \int_0^1 \frac{dt}{\sqrt{\left(1-t^2\right)\left(1-z t^2\right)}}\,,$  $ \mathrm{E}(z) \equiv  \int_0^1 dt\, \frac{\sqrt{1-z t^2}}{\sqrt{1-t^2}}, $
 and 
\begin{eqnarray} 
&& h_{1}= \left(-3+377 \left(1+|p_\infty^2|\right)-1017 \left(1+|p_\infty^2|\right)^2+515 \left(1+|p_\infty^2|\right)^3\right), \nonumber \\
&& h_{2}= \left(169+380
\left(1+|p_\infty^2|\right)\right), \nonumber \\
&& h_{3}= \left(834+2095 \sqrt{1+|p_\infty^2|}+1200 \left(1+|p_\infty^2|\right)\right), \nonumber \\
&& h_{4}= \left(1183+2929 \sqrt{1+|p_\infty^2|}+2660
\left(1+|p_\infty^2|\right)+1200 \left(1+|p_\infty^2|\right)^{3/2}\right), \nonumber \\
&& h_{5}= \left(-12+76 \sqrt{1+|p_\infty^2|}-129 \left(1+|p_\infty^2|\right)+60 \left(1+|p_\infty^2|\right)^{3/2}+30
\left(1+|p_\infty^2|\right)^2\right.\nonumber\\ &&\left.\ \ \ -25 \left(1+|p_\infty^2|\right)^3\right), \nonumber \\
&& h_{6}= \left(1151-3336 \sqrt{1+|p_\infty^2|}+3148 \left(1+|p_\infty^2|\right)-912 \left(1+|p_\infty^2|\right)^{3/2}+339
\left(1+|p_\infty^2|\right)^2\right.\nonumber\\ &&\left.\ \ \ -552 \left(1+|p_\infty^2|\right)^{5/2}+210 \left(1+|p_\infty^2|\right)^3\right), \nonumber \\ 
&& h_{7}=- \sqrt{1+|p_\infty^2|} \left(-3+2 \left(1+|p_\infty^2|\right)\right)
\left(4-15 \sqrt{1+|p_\infty^2|}+15 \left(1+|p_\infty^2|\right)\right), \nonumber \\
&& h_{8}= \left(-1049-496 \sqrt{1+|p_\infty^2|}+8700 \left(1+|p_\infty^2|\right)-16658 \left(1+|p_\infty^2|\right)^{3/2}\right.\nonumber\\ &&\left.\ \ \ +9563
\left(1+|p_\infty^2|\right)^2+13176 \left(1+|p_\infty^2|\right)^{5/2}-15822 \left(1+|p_\infty^2|\right)^3\right.\nonumber\\ &&\left.  \ \ \  -1338 \left(1+|p_\infty^2|\right)^{7/2}+3456 \left(1+|p_\infty^2|\right)^4+420 \left(1+|p_\infty^2|\right)^{9/2}\right), \nonumber \\ 
&& h_{9}=
\left(-210(1+ \sqrt{1+|p_\infty^2|})+885 \left(1+|p_\infty^2|\right)+885 \left(1+|p_\infty^2|\right)^{3/2}-3457 \left(1+|p_\infty^2|\right)^2\right.\nonumber\\ &&\left.  \ \ \ -3457 \left(1+|p_\infty^2|\right)^{5/2}+9593 \left(1+|p_\infty^2|\right)^3+9593
\left(1+|p_\infty^2|\right)^{7/2}+3259 \left(1+|p_\infty^2|\right)^4 \right.\nonumber\\ &&\left.  \ \ \ -181493 \left(1+|p_\infty^2|\right)^{9/2}+535259 \left(1+|p_\infty^2|\right)^5-500785 \left(1+|p_\infty^2|\right)^{11/2}\right.\nonumber\\ &&\left.  \ \ \  -32675
\left(1+|p_\infty^2|\right)^6+333545 \left(1+|p_\infty^2|\right)^{13/2}-304761 \left(1+|p_\infty^2|\right)^7\right.\nonumber\\ &&\left.  \ \ \  +232751 \left(1+|p_\infty^2|\right)^{15/2}+74431 \left(1+|p_\infty^2|\right)^8-216185
\left(1+|p_\infty^2|\right)^{17/2}  -34080 \left(1+|p_\infty^2|\right)^9\right.\nonumber\\ &&\left.  \ \ \ +116100 \left(1+|p_\infty^2|\right)^{19/2}+11340 \left(1+|p_\infty^2|\right)^{10}-22680 \left(1+|p_\infty^2|\right)^{21/2}\right), \nonumber \\
&& h_{10}=
\left(-129+366 \sqrt{1+|p_\infty^2|}+444 \left(1+|p_\infty^2|\right)-1432 \left(1+|p_\infty^2|\right)^{3/2}+27 \left(1+|p_\infty^2|\right)^2\right.\nonumber\\ &&\left.  \ \ \ \ \ +970 \left(1+|p_\infty^2|\right)^{5/2}+50 \left(1+|p_\infty^2|\right)^3-280
\left(1+|p_\infty^2|\right)^{7/2}\right), \nonumber \\
&& h_{11}= \left(2074+10643 \sqrt{1+|p_\infty^2|}+18958 \left(1+|p_\infty^2|\right)+11391 \left(1+|p_\infty^2|\right)^{3/2}\right.\nonumber\\ &&\left.  \ \ \ \ \ +5242 \left(1+|p_\infty^2|\right)^2-9826
\left(1+|p_\infty^2|\right)^{5/2}+1818 \left(1+|p_\infty^2|\right)^3+13198 \left(1+|p_\infty^2|\right)^{7/2}\right.\nonumber\\ &&\left.  \ \ \ \ \  -700 \left(1+|p_\infty^2|\right)^4-10065 \left(1+|p_\infty^2|\right)^{9/2}+2835 \left(1+|p_\infty^2|\right)^{11/2}\right), \nonumber \\ 
&& h_{12}=
\sqrt{1+|p_\infty^2|} \left(5369+8077 \left(1+|p_\infty^2|\right)-5014 \left(1+|p_\infty^2|\right)^2+4874 \left(1+|p_\infty^2|\right)^3\right.\nonumber\\ &&\left.  \ \ \ \ \ -2955 \left(1+|p_\infty^2|\right)^4+945 \left(1+|p_\infty^2|\right)^5\right), \nonumber \\
&& h_{13}=
\sqrt{1+|p_\infty^2|} \left(-1965+2169 \sqrt{1+|p_\infty^2|}+1289 \left(1+|p_\infty^2|\right)-2211 \left(1+|p_\infty^2|\right)^{3/2}\right.\nonumber\\ &&\left.  \ \ \ \ \ -856 \left(1+|p_\infty^2|\right)^2+90 \left(1+|p_\infty^2|\right)^{5/2}+580
\left(1+|p_\infty^2|\right)^3+280 \left(1+|p_\infty^2|\right)^{7/2}\right), \nonumber \\
&& h_{14}= \sqrt{1+|p_\infty^2|} \left(-3+2 \left(1+|p_\infty^2|\right)\right) \left(85-82 \sqrt{1+|p_\infty^2|}-716
\left(1+|p_\infty^2|\right)\right.\nonumber\\ &&\left.  \ \ \ \ \ +380 \left(1+|p_\infty^2|\right)^{3/2}+1537 \left(1+|p_\infty^2|\right)^2-610 \left(1+|p_\infty^2|\right)^{5/2}-890 \left(1+|p_\infty^2|\right)^3\right.\nonumber\\ &&\left.  \ \ \ \ \ +280 \left(1+|p_\infty^2|\right)^{7/2}\right), \nonumber \\ 
&& h_{15}=
\left(-5+76 \sqrt{1+|p_\infty^2|}-150 \left(1+|p_\infty^2|\right)+60 \left(1+|p_\infty^2|\right)^{3/2}+35 \left(1+|p_\infty^2|\right)^2\right), \nonumber \\
&& h_{16}= \sqrt{1+|p_\infty^2|} \left(-3+2 \left(1+|p_\infty^2|\right)\right)
\left(11-30 \left(1+|p_\infty^2|\right)+35 \left(1+|p_\infty^2|\right)^2\right), \nonumber \\
&& h_{17}= \left(299-1216 \sqrt{1+|p_\infty^2|}+1732 \left(1+|p_\infty^2|\right)-960 \left(1+|p_\infty^2|\right)^{3/2}\right.\nonumber\\ &&\left.  \ \ \ \ \ +690
\left(1+|p_\infty^2|\right)^2-860 \left(1+|p_\infty^2|\right)^3+315 \left(1+|p_\infty^2|\right)^4\right), \nonumber \\
&& h_{18}= \left(21+65 \left(1+|p_\infty^2|\right)-145 \left(1+|p_\infty^2|\right)^2+315
\left(1+|p_\infty^2|\right)^3\right), \nonumber \\
&& h_{19}= \left(102-4983 \sqrt{1+|p_\infty^2|}+12882 \left(1+|p_\infty^2|\right)-11744 \left(1+|p_\infty^2|\right)^{3/2}\right.\nonumber\\ &&\left.  \ \ \ \ \ -2154 \left(1+|p_\infty^2|\right)^2+20405
\left(1+|p_\infty^2|\right)^{5/2}-17562 \left(1+|p_\infty^2|\right)^3\right.\nonumber\\ &&\left.  \ \ \ \ \ +234 \left(1+|p_\infty^2|\right)^{7/2}+1932 \left(1+|p_\infty^2|\right)^4+840 \left(1+|p_\infty^2|\right)^{9/2}\right), \nonumber \\
&& h_{20}=
\left(-45+207 \left(1+|p_\infty^2|\right)-1471 \left(1+|p_\infty^2|\right)^2+13349 \left(1+|p_\infty^2|\right)^3\right.\nonumber\\ &&\left.  \ \ \ \ \ -37478 \left(1+|p_\infty^2|\right)^{7/2}+63848 \left(1+|p_\infty^2|\right)^4-47540
\left(1+|p_\infty^2|\right)^{9/2}\right.\nonumber\\ &&\left.  \ \ \ \ \ -9872 \left(1+|p_\infty^2|\right)^5+16138 \left(1+|p_\infty^2|\right)^{11/2}+14128 \left(1+|p_\infty^2|\right)^6\right.\nonumber\\ &&\left.  \ \ \ \ \ +7824 \left(1+|p_\infty^2|\right)^{13/2}-23840
\left(1+|p_\infty^2|\right)^7+4320 \left(1+|p_\infty^2|\right)^{15/2}\right.\nonumber\\ &&\left.  \ \ \ \ \ +3600 \left(1+|p_\infty^2|\right)^8\right), \nonumber \\&& 
h_{21}= \left(-124+285 \sqrt{1+|p_\infty^2|}+660 \left(1+|p_\infty^2|\right)-1480
\left(1+|p_\infty^2|\right)^{3/2}\right.\nonumber\\ &&\left.  \ \ \ \ \ -400 \left(1+|p_\infty^2|\right)^2+1425 \left(1+|p_\infty^2|\right)^{5/2}-350 \left(1+|p_\infty^2|\right)^{7/2}\right), \nonumber \\
&& h_{22}= \left(1759+6744
\sqrt{1+|p_\infty^2|}+3692 \left(1+|p_\infty^2|\right)+2044 \left(1+|p_\infty^2|\right)^{3/2}\right.\nonumber\\ &&\left.  \ \ \ \ \ +2787 \left(1+|p_\infty^2|\right)^2+1112 \left(1+|p_\infty^2|\right)^{5/2}+210 \left(1+|p_\infty^2|\right)^3-300
\left(1+|p_\infty^2|\right)^{7/2}\right), \nonumber \\
&& h_{23}= \sqrt{1+|p_\infty^2|} \left(-852-283 \left(1+|p_\infty^2|\right)-140 \left(1+|p_\infty^2|\right)^2+75 \left(1+|p_\infty^2|\right)^3\right), \nonumber \\
&& h_{24}=
\sqrt{1+|p_\infty^2|} \left(-3+2 \left(1+|p_\infty^2|\right)\right) \left(1151-3504 \sqrt{1+|p_\infty^2|}+3148 \left(1+|p_\infty^2|\right)\right.\nonumber\\ &&\left.  \ \ \ \ \ -576 \left(1+|p_\infty^2|\right)^{3/2}+339 \left(1+|p_\infty^2|\right)^2-720
\left(1+|p_\infty^2|\right)^{5/2}+210 \left(1+|p_\infty^2|\right)^3\right), \nonumber \\
&& h_{25}= \sqrt{1+|p_\infty^2|} \left(-3+2 \left(1+|p_\infty^2|\right)\right) \left(96-93 \sqrt{1+|p_\infty^2|}-768
\left(1+|p_\infty^2|\right)\right.\nonumber\\ &&\left.  \ \ \ \ \ +432 \left(1+|p_\infty^2|\right)^{3/2}+1632 \left(1+|p_\infty^2|\right)^2-705 \left(1+|p_\infty^2|\right)^{5/2}\right.\nonumber\\ &&\left.  \ \ \ \ \ -960 \left(1+|p_\infty^2|\right)^3+350 \left(1+|p_\infty^2|\right)^{7/2}\right), \nonumber \\&& h_{26}=
\left(1+|p_\infty^2|\right) \left(3-2 \left(1+|p_\infty^2|\right)\right)^2 \left(11-30 \left(1+|p_\infty^2|\right)+35 \left(1+|p_\infty^2|\right)^2\right), \nonumber \\
&& h_{27}= \left(8+19 \sqrt{1+|p_\infty^2|}+60
\left(1+|p_\infty^2|\right)+15 \left(1+|p_\infty^2|\right)^{3/2}\right), \nonumber \\
&& h_{28}= \sqrt{1+|p_\infty^2|} \left(63+768 \left(1+|p_\infty^2|\right)-645 \left(1+|p_\infty^2|\right)^2+70 \left(1+|p_\infty^2|\right)^3\right), \nonumber \\
&& h_{29}=
\left(60+333 \left(1+|p_\infty^2|\right)+90 \left(1+|p_\infty^2|\right)^2-75 \left(1+|p_\infty^2|\right)^3\right), \nonumber \\
&& h_{30}= \left(12+76 \sqrt{1+|p_\infty^2|}+129 \left(1+|p_\infty^2|\right)+60
\left(1+|p_\infty^2|\right)^{3/2}-30 \left(1+|p_\infty^2|\right)^2\right.\nonumber\\ &&\left.  \ \ \ \ \ +25 \left(1+|p_\infty^2|\right)^3\right),\nonumber \\ 
&& h_{61}= 35 (\sqrt{1+|p_\infty^2| }-1 ) (1+\sqrt{1+|p_\infty^2| }) \Big(1-18 (1+|p_\infty^2| )+33 (1+|p_\infty^2| )^2\Big), \nonumber \\ 
&& h_{62}=
(-45+207 (1+|p_\infty^2| )-1471 (1+|p_\infty^2| )^2+13349 (1+|p_\infty^2| )^3 \nonumber\\ &&  \ \ \ \ \ -38135 (1+|p_\infty^2| )^{7/2}+64424 (1+|p_\infty^2| )^4-32177
(1+|p_\infty^2| )^{9/2}-15056 (1+|p_\infty^2| )^5\nonumber\\ &&  \ \ \ \ \ -25145 (1+|p_\infty^2| )^{11/2}+27952 (1+|p_\infty^2| )^6+33249 (1+|p_\infty^2| )^{13/2}-35360
(1+|p_\infty^2| )^7\nonumber\\ &&  \ \ \ \ \  +4320 (1+|p_\infty^2| )^{15/2}+3600 (1+|p_\infty^2| )^8), \nonumber \\ 
&& h_{63}= (1+|p_\infty^2| ) (2 (1+|p_\infty^2| )-3)
(2 (1+|p_\infty^2| )-1) (11-30 (1+|p_\infty^2| )+35 (1+|p_\infty^2| )^2), \nonumber \\ 
&& h_{64}= (2681 \sqrt{1+|p_\infty^2| }-102-6210
(1+|p_\infty^2| )+10052 (1+|p_\infty^2| )^{3/2}-9366 (1+|p_\infty^2| )^2\nonumber\\ &&  \ \ \ \ \ -8491 (1+|p_\infty^2| )^{5/2}+15018 (1+|p_\infty^2| )^3+702 (1+|p_\infty^2| )^{7/2}-4140
(1+|p_\infty^2| )^4), \nonumber \\ 
&& h_{65}= (124-295 \sqrt{1+|p_\infty^2| }-508 (1+|p_\infty^2| )+1200 (1+|p_\infty^2| )^{3/2}+216 (1+|p_\infty^2| )^2\nonumber\\ &&  \ \ \ \ \ -755
(1+|p_\infty^2| )^{5/2}-240 (1+|p_\infty^2| )^3+210 (1+|p_\infty^2| )^{7/2}), \nonumber \\ 
&& h_{66}= \sqrt{1+|p_\infty^2| } (2 (1+|p_\infty^2| )-3)
(2 (1+|p_\infty^2| )-1) (11-30 (1+|p_\infty^2| )+35 (1+|p_\infty^2| )^2), \nonumber \\ 
&& h_{67}= (1-\sqrt{1+|p_\infty^2| }) (-947-3177
\sqrt{1+|p_\infty^2| }+14910 (1+|p_\infty^2| )-26710 (1+|p_\infty^2| )^{3/2} \nonumber\\ &&  \ \ \ \ \  +18929 (1+|p_\infty^2| )^2+21667 (1+|p_\infty^2| )^{5/2}-30840 (1+|p_\infty^2| )^3-2040
(1+|p_\infty^2| )^{7/2} \nonumber\\ &&  \ \ \ \ \  +7596 (1+|p_\infty^2| )^4+420 (1+|p_\infty^2| )^{9/2}), \nonumber \\ 
&& h_{68}= (\sqrt{1+|p_\infty^2| }-1 ) (253-661 \sqrt{1+|p_\infty^2| }-952
(1+|p_\infty^2| )+2632 (1+|p_\infty^2| )^{3/2}\nonumber\\ &&  \ \ \ \ \  +189 (1+|p_\infty^2| )^2-1725 (1+|p_\infty^2| )^{5/2}-290 (1+|p_\infty^2| )^3+490 (1+|p_\infty^2| )^{7/2}),
\end{eqnarray}

\section {The coefficients $a_i$ in the metric (\ref{Mmetric})}\label{Tp}

The coefficients $a_i$ in the metric (\ref{Mmetric}) are given by the following expressions 
\begin{eqnarray} \label{parameters1}
a_2&=&\frac{3 (\Gamma-1)\left(4+5 |p_\infty^2|\right) }{\left(2+3 |p_\infty^2|\right) \Gamma },\nonumber \\
a_3&=&\frac{3 }{2 \left(3+4 |p_\infty^2|\right)}\Bigg\{\frac{108+3
|p_\infty^2| \left(85+|p_\infty^2| (50-32 \Gamma )-56 \Gamma \right)-74 \Gamma }{\left(2+3 |p_\infty^2|\right) \Gamma }-\frac{2 \,T_3^p}{\sqrt{|p_\infty^2|}}\nonumber \\ &-&\frac{1}{
\Gamma ^2}\Big[17+18 |p_\infty^2|+2\nu \Big(1-\frac{103 \sqrt{1+|p_\infty^2|}}{3}-18 \left(1+|p_\infty^2|\right)-\frac{2}{3} \left(1+|p_\infty^2|\right)^{3/2}\nonumber \\ &+&\frac{3 (1+2
|p_\infty^2|) (4+5 |p_\infty^2|) \Gamma }{(1+\sqrt{1+|p_\infty^2|}) (1+\Gamma )}  +\frac{8 (11+4 |p_\infty^2|-4 |p_\infty^2|^2) }{\sqrt{|p_\infty^2|}}  \text{arcsinh}\sqrt{\frac{\sqrt{1+|p_\infty^2|}-1}{2}}\Big)\Big] \Bigg\},\nonumber \\
a_4&=&  \frac{\Gamma-1
}{4  (4+5 |p_\infty^2|) \Gamma ^3}\Big\{\frac{560}{|p_\infty^2|}+16 \Big[105+8 \Gamma  (1+\Gamma )\Big]+3 |p_\infty^2|  \Big[385+43 \Gamma  (1+\Gamma )\Big]\Big\}\nonumber \\
&+& \frac{4}{ (4+5 |p_\infty^2|)} \Bigg\{\frac{|p_\infty^2| }{48}
 \left(-390 a_2+51 a_2^2-164 a_3\right)+ \Big[(a_2-8) a_2-\frac{10\, a_3}{3}\Big]\nonumber \\ &-&\frac{T_4^{\nu}}{|p_\infty^2|}+\frac{2 \left(1+2 |p_\infty^2|\right) T_3^p}{|p_\infty^2|^{3/2}}-\frac{8\, T_4^p}{3 \pi |p_\infty^2| }\Bigg\}, 
     \end{eqnarray}
where      
 \begin{eqnarray} 
 T_3^p&=& \frac{2 (1+2  |p_\infty^2|)^2 \nu }{3\,
 |p_\infty^2| \,\Gamma ^2} \Big( 3-5  |p_\infty^2|+\frac{6 \sqrt{1+ |p_\infty^2|} (2  |p_\infty^2|-1) }{\sqrt{ |p_\infty^2|}} \text{arcsinh}\sqrt{\frac{\sqrt{1+|p_\infty^2|}-1}{2}} \Big),\nonumber \\ 
 T_4^p&=&\frac{\pi  \nu }{96  |p_\infty^2|^2 \Gamma ^5} \Big[-3\, \text{arccosh}[\sqrt{1+ |p_\infty^2|}] \Big[(1+\sqrt{1+ |p_\infty^2|}) \nu  h_{14}+\Big(1+(\sqrt{1+ |p_\infty^2|}\nonumber \\ 
&-&3)
\nu \Big) h_{25}\Big]+12 \,\text{arcsinh}\sqrt{\frac{\sqrt{1+|p_\infty^2|}-1}{2}} \,\Big(h_{63}+\frac{ |p_\infty^2|^2 \nu  h_{66}}{2+ |p_\infty^2|+2 \sqrt{1+ |p_\infty^2|}}\Big)\nonumber \\ 
&+&\sqrt{ |p_\infty^2|}
\Big(h_{64}+\nu  h_{67}+6  |p_\infty^2| (h_{65}+\nu  h_{68}) \text{log}[\frac{1}{2} (1+\sqrt{1+ |p_\infty^2|})]\Big)\Big]\nonumber \\
 &+&\frac{ |p_\infty^2|^2 \pi  \nu ^2 }{1536 \Gamma ^5}\Bigg[\frac{4 (1+ |p_\infty^2|) (1+\sqrt{1+ |p_\infty^2|}) h_{20}-h_{9}}{ |p_\infty^2|^3 (1+ |p_\infty^2|)^{9/2}}-\frac{72 \text{arccosh}[\sqrt{1+ |p_\infty^2|}]^2
h_{26}}{ |p_\infty^2|^3 (\sqrt{1+ |p_\infty^2|}-1)}\nonumber \\ 
&+&\frac{24 (8  |p_\infty^2| h_{23}-h_{12}) \text{log}(1+ |p_\infty^2|)}{ |p_\infty^2|^2 (\sqrt{1+ |p_\infty^2|}-1)}+\frac{24
(h_{11}+2  |p_\infty^2| h_{22}) \text{log}[\frac{1}{2} (1+\sqrt{1+ |p_\infty^2|})]}{ |p_\infty^2|^2 (\sqrt{1+ |p_\infty^2|}-1)}\nonumber \\ 
&-&\frac{288 (h_{15}-4
h_{27}) \text{log}[\frac{1}{2} (1+\sqrt{1+ |p_\infty^2|})]^2}{\sqrt{1+ |p_\infty^2|}-1}
\nonumber \\ 
&+&\frac{48\, \text{arccosh}(\sqrt{1+ |p_\infty^2|}) }{ |p_\infty^2|^3
\sqrt{ |p_\infty^2|} (\sqrt{1+ |p_\infty^2|}-1)} \Big(2
\Big( |p_\infty^2| (\sqrt{1+ |p_\infty^2|}-3)+4 (\sqrt{1+ |p_\infty^2|}-1)\Big) h_{13}\nonumber \\ &
-& |p_\infty^2| h_{24}+3  |p_\infty^2|^2 (h_{16}+h_{28}) \text{log}[\frac{1}{2} (1+\sqrt{1+ |p_\infty^2|})]\Big)
\nonumber \\ 
&+&\frac{24 (3  |p_\infty^2| h_{18}+8 h_{29}) \text{Li}_2\Big[\frac{1-\sqrt{1+ |p_\infty^2|}}{1+\sqrt{1+ |p_\infty^2|}}\Big]}{\sqrt{1+ |p_\infty^2|}-1}+\frac{36
(h_{17}+8 h_{30}) \text{Li}_2\Big[\frac{\sqrt{1+ |p_\infty^2|}-1}{1+\sqrt{1+ |p_\infty^2|}}\Big]}{\sqrt{1+ |p_\infty^2|}-1}\Bigg],\label{xrr} \\ 
 T_{4}^{\nu}&=&\frac{|p_\infty^2|^2 \nu }{144
\Gamma ^3} \Bigg\{\frac{36 }{|p_\infty^2|^2} \text{ E}\Big(\frac{\sqrt{1+|p_\infty^2|}-1}{1+\sqrt{1+|p_\infty^2|}}\Big)
\text{ K}\Big(\frac{\sqrt{1+|p_\infty^2|}-1}{1+\sqrt{1+|p_\infty^2|}}\Big) h_{4}-\frac{126 \text{ E}\Big(\frac{\sqrt{1+|p_\infty^2|}-1}{1+\sqrt{1+|p_\infty^2|}}\Big)^2 h_{2}}{|p_\infty^2| (\sqrt{1+|p_\infty^2|}-1)}\nonumber \\ 
&-&\frac{36}{|p_\infty^2|^2}\text{ K}\Big(\frac{\sqrt{1+|p_\infty^2|}-1}{1+\sqrt{1+|p_\infty^2|}}\Big)^2 h_{3}
-\frac{24 \,\pi^2\, h_{5}}{|p_\infty^2|}+\frac{12\, h_{24} \, \text{arccosh}\sqrt{1+|p_\infty^2|}
}{|p_\infty^2|^{7/2}}\nonumber \\ 
&+&\frac{18 \,h_{26}\,\text{arccosh}(\sqrt{1+|p_\infty^2|})^2 }{|p_\infty^2|^4}-\frac{h_{62}}{|p_\infty^2|^3 (1+|p_\infty^2|)^{7/2}}-\frac{48\,
h_{23}\, \text{log}(1+|p_\infty^2|)}{|p_\infty^2|^2}\nonumber \\ 
&-&\Big(\frac{12\, h_{6} }{|p_\infty^2|^2}+\frac{36\, h_{16}\,\text{arccosh}(\sqrt{1+|p_\infty^2|})
 }{|p_\infty^2|^{5/2}}\Big)\text{log}\Big(\frac{1}{2} \Big(\sqrt{1+|p_\infty^2|}-1\Big)\Big)\nonumber \\ 
&-&\Big(\frac{12\, h_{22} }{|p_\infty^2|^2}+\frac{36\, h_{28} \, \text{arccosh}\Big(\sqrt{1+|p_\infty^2|}\Big)}{|p_\infty^2|^{5/2}}\Big)\text{log}\Big(\frac{1}{2} \Big(1+\sqrt{1+|p_\infty^2|}\Big)\Big)\nonumber \\ 
&+&\frac{72 h_{15}
\text{log}(\frac{1}{2} (\sqrt{1+|p_\infty^2|}-1)) \text{log}(\frac{1}{2} (1+\sqrt{1+|p_\infty^2|}))}{|p_\infty^2|}-\frac{48 h_{29} }{|p_\infty^2|} \text{Li}_2(\frac{1-\sqrt{1+|p_\infty^2|}}{1+\sqrt{1+|p_\infty^2|}})\nonumber \\
&-&\frac{288\, h_{27}\,
\text{log}\Big(\frac{1}{2} \Big(1+\sqrt{1+|p_\infty^2|}\Big)\Big)^2}{|p_\infty^2|}-\frac{576\, h_{7} \, \sqrt{|p_\infty^2|} \, \text{Li}_2\Big(\sqrt{\frac{\sqrt{1+|p_\infty^2|}-1}{1+\sqrt{1+|p_\infty^2|}}}\Big)}{|p_\infty^2|^2 (1+\sqrt{1+|p_\infty^2|})}
\nonumber \\
&+&\frac{72\, (2\, \sqrt{|p_\infty^2|}\,
h_{7}-|p_\infty^2| (1+\sqrt{1+|p_\infty^2|})\, h_{30})  \text{Li}_2\Big(\frac{\sqrt{1+|p_\infty^2|}-1}{1+\sqrt{1+|p_\infty^2|}}\Big)}{|p_\infty^2|^2 (1+\sqrt{1+|p_\infty^2|})}\Bigg\}\nonumber \\ 
&-&\frac{2 (1+2 |p_\infty^2|) \nu }{\Gamma
^2} \Bigg[1-\frac{103 \sqrt{1+|p_\infty^2|}}{3}-18 (1+|p_\infty^2|)-\frac{2}{3} (1+|p_\infty^2|)^{3/2}\nonumber \\ &+&\frac{3 (1+2
|p_\infty^2|) (4+5 |p_\infty^2|) \Gamma }{(1+\sqrt{1+|p_\infty^2|}) (1+\Gamma )}+\frac{8 (11+4 |p_\infty^2|-4 |p_\infty^2|^2) \text{arcsinh}\sqrt{\frac{\sqrt{1+|p_\infty^2|}-1}{2}} }{\sqrt{|p_\infty^2|}}\Bigg].
 \label{f4}
 \end{eqnarray}

 \section{\label{APPC} Existence of a null tetrad that satisfies the specified gauge condition }
  
We now demonstrate the existence of a null tetrad that satisfies the specified gauge condition. 
In the original null tetrad $(l_{\mu},\,n_{\mu},\, m_{\mu},\, \bar{m}_{\mu})$, if  
$${\cal{G}}_4^B\neq0,$$ 
we can perform a rotation of Class I  \cite{Chandrasekhar}
\begin{align}
(l'_{\mu},\,n'_{\mu},\, m'_{\mu},\, \bar{m}'_{\mu})^T=\begin{pmatrix}
			1 & 0 & 0 & 0 \\
			o o^{*} & 1 & o^{*} & o \\
			o & 0 & 1 & 0 \\
			o^{*} & 0 & 0 & 1 \\
		\end{pmatrix}(l_{\mu},\,n_{\mu},\, m_{\mu},\, \bar{m}_{\mu})^T,
		\end{align}
where $o$ is an infinitesimal complex function. 
Notably, neglecting higher-order infinitesimals,  in Eq. \eqref{eq06J1}, the quantities affected by the infinitesimal transformations of Class I are
 	\begin{align}
		\nonumber
		\nonumber T_{4}'^{B} =
		& \nonumber T_{4}^{B} + 2 \left( \varDelta + 3 \gamma - \bar{\gamma} + \bar{\mu} + 4 \mu\right) (\bar{\delta} + 2 \alpha - 2 \bar{\tau} ) \Phi_{11}\, o^* 
		\\
		& +2 \left( \bar{\delta} + 3 \alpha + \bar{\beta} - \bar{\tau} + 4 \pi \right) (\varDelta  + 2 \bar{\mu}+ 2 \gamma) \Phi_{11}\, o^* ,\label{GG1}
		\\
		\nonumber
		{\cal{G}}'^B_4=
		& {\cal{G}}_4^B 
		+\Big[({\bf{\Delta}}+3\gamma-\bar{\gamma}+4\mu+\bar{\mu})(\bar{\delta}+2\alpha+4\pi) 
-(\bar{\delta}+3\alpha+\bar{\beta}-\bar{\tau}+4\pi)\cdot \nonumber \\ &({\bf{\Delta}}+2\gamma+4\mu)\Big] \psi_{2}\, o^*		
		+ 4 \left( \varDelta + 3 \gamma - \bar{\gamma} + 2 \bar{\mu} + 2 \mu \right) ( \Phi_{11}  (\bar{\delta} + 2 \alpha + \pi ) \,o^*) \nonumber 
		\\
		& + 4 \left( \bar{\delta} + \pi - \bar{\tau} \right)  \Phi_{11}  (\varDelta + 2 \gamma + \mu) \, o^* .\label{GG2}
	\end{align}
Eqs. (\ref{GG1}) and (\ref{GG2}) pertain exclusively to a single function $ o^{*} $, with the exception of $ T_4^B $ and $ G_4^B $. Thus, we can identify a null tetrad that meets the specified gauge condition by selecting an appropriate function $ o^{*} $ that satisfies the following equation
\begin{align}
		& {\cal{G}}_4^B 
		+\Big[({\bf{\Delta}}+3\gamma-\bar{\gamma}+4\mu+\bar{\mu})(\bar{\delta}+2\alpha+4\pi) 
-(\bar{\delta}+3\alpha+\bar{\beta}-\bar{\tau}+4\pi)\cdot \nonumber \\ &({\bf{\Delta}}+2\gamma+4\mu)\Big] \psi_{2}\, o^*		
		+4 \left( \varDelta + 3 \gamma - \bar{\gamma} + 2 \bar{\mu} + 2 \mu \right) (\Phi_{11}  (\bar{\delta} + 2 \alpha + \pi ) \,o^*) \nonumber 
		\\
		& + 4 \left( \bar{\delta} + \pi - \bar{\tau} \right) \Phi_{11}  (\varDelta + 2 \gamma + \mu) \, o^* + 2 \left( \varDelta + 3 \gamma - \bar{\gamma} + \bar{\mu} + 4 \mu\right) (\bar{\delta} + 2 \alpha - 2 \bar{\tau} ) \Phi_{11}\, o^*\nonumber \\
&+2 \left( \bar{\delta} + 3 \alpha + \bar{\beta} - \bar{\tau} + 4 \pi \right) (\varDelta  + 2 \bar{\mu}+ 2 \gamma) \Phi_{11}\, o^*=0.		
		 \label{GG3}
	\end{align}	

\section{The coefficients $A_{ij}$ in Eq. \ref{NTgenTTsl}} \label{AijA}
\begin{align}
A _{nn\,0}&=-\frac{2 r^3}{ \sqrt{2\pi}\,\Delta ^2 }\,
\,
\mathscr{L}_1^{\dag}\Big[r^4\mathscr{L}_2^{\dag}\Big(\frac{1}{r^3}\,_{-2}Y_{\ell m}(\theta)\Big)  \Big],\nonumber \\
A _{{\bar m}n\,0}&=\frac{2 }{ \sqrt{\pi}\Delta   } 
\,   r^3 \Big[\Big(\frac{i K}{ \Delta   }+\frac{2}{r}\Big) \mathscr{L}_2^{\dag}\Big]\,_{-2}Y_{\ell m}(\theta) ,\nonumber \\
A _{{\bar m}{\bar m}\,0}
&=\frac{r^2 }{ \sqrt{2\pi}}\,  
 \Bigl[
i\Bigl(\frac{K }{ \Delta   }\Bigr)' +\frac{K^2 }{ \Delta   ^2}\nonumber - 2 i\frac{ K }{ \Delta   r}\Bigr]\,_{-2}Y_{\ell m}(\theta) ,\nonumber \\
A _{{\bar m}n\,1}&=\frac{
 2 r^3}{ \sqrt{\pi}\Delta    }\,
 \Big[ \mathscr{L}_2^{\dag}\Big]\,_{-2}Y_{\ell m}(\theta) 
,\nonumber \\
A _{{\bar m}{\bar m}\,1}
&=-\frac{ 2  r^2 }{ \sqrt{2\pi} }
\,
\Bigl(  \frac{i K}{ \Delta   }+\frac{1}{r}\Bigr)\,_{-2}Y_{\ell m}(\theta)  ,\nonumber \\
A _{{\bar m}{\bar m}\,2}
&=-\frac{r^2}{ \sqrt{2\pi }}\,
\,_{-2}Y_{\ell m}(\theta).
\label{Aijsk} 
\end{align}


\bibliography{mybib}

\end{document}